\documentclass[preprint]{aastex}
\usepackage{epsfig}

\newcommand{\nci}{115}
\newcommand{\ncii}{993}
\newcommand{\ici}{260}
\newcommand{\icii}{915}
\newcommand{\allci}{375}
\newcommand{\allcii}{1908}
\newcommand{\allyso}{2283}
\newcommand{\mci}{800} 
\newcommand{\mcii}{1200}

\newcommand{\numimfci}{1600}
\newcommand{\numimfcii}{2400}

\newcommand{\sigsfrmed}{8.2}

\newcommand{\mallysolow}{4800}
\newcommand{\mallysohigh}{17400}
\newcommand{\unitssigsfr}{$M_{\odot}$ Myr$^{-1}$ pc$^{-2}$}

\slugcomment{Accepted to the Astrophysical Journal}
\shorttitle{NGC 6334}
\shortauthors{Willis et al.}
\title{A Wide-field Near- and Mid-Infrared Census of Young Stars in NGC 6334}
\begin{document}


\author{S. Willis\altaffilmark{3}, M. Marengo}
\affil{Department of Physics and Astronomy, Iowa State University, Ames, IA 50010, USA}
\author{L. Allen}
\affil{National Optical Astronomy Observatories, Tucson, AZ 85719, USA}
\author{G.G. Fazio, and H. A. Smith}
\affil{Harvard-Smithsonian Center for Astrophysics, Cambridge, MA 02138, USA}
\author{S. Carey}
\affil{Spitzer Science Center, California Institute of Technology, Pasadena, CA 91125, USA}

\email{swillis@cfa.harvard.edu}
\begin{abstract}
This paper presents a study of the rate and efficiency of star formation in the NGC 6334 star forming region. We obtained observations at $J$, $H$, and $K_{s}$ taken with the NOAO Extremely Wide-Field Infrared Imager (NEWFIRM) and combined them with observations taken with the Infrared Array Camera (IRAC) camera on the Spitzer Space Telescope at wavelengths λ = 3.6, 4.5, 5.8, and 8.0 $\mu$m. We also analyzed previous observations taken at 24 $\mu$m using the Spitzer MIPS camera as part of the MIPSGAL survey. We have produced a point source catalog with $>700,000$ entries. We have identified \allyso{} YSO candidates, \allci{} Class I YSOs and \allcii{} Class II YSOs using a combination of existing IRAC-based color classification schemes that we have extended and validated to the near-IR for use with warm Spitzer data. We have identified multiple new sites of ongoing star formation activity along filamentary structures extending tens of parsecs beyond the central molecular ridge of NGC 6334. By mapping the extinction we derived an estimate for the gas mass, 2.2$\times10^{5}M_{\odot}$. The heavy concentration of protostars along the dense filamentary structures indicates that NGC 6334 may be undergoing a ``mini-starburst'' event with $\Sigma_{SFR}>\sigsfrmed{}$ \unitssigsfr{} and $SFE>0.10$. We have used these estimates to place NGC 6334 in the Kennicutt-Schmidt diagram to help bridge the gap between observations of local low-mass star forming regions and star formation in other galaxies.

\end{abstract}

\section{INTRODUCTION}\label{sec:Introduction}
Active star formation in other galaxies can be inferred from several indicators of the presence of short-lived massive stars, including strong UV flux and H$\alpha$ emission from excited hydrogen gas. Although these massive stars dominate the observable emission, they represent only a tiny fraction of the total stellar mass being formed; the majority of this is contained in much more numerous low mass stars. Empirical relations, such as the Kennicutt-Schmidt law \citep{ken98} have matched the observed surface density of total rate of star formation to the surface density of gas in other galaxies. Recent detailed studies of nearby Galactic star forming regions obtained a complete census of the young stellar content and found star formation rates orders of magnitude higher than predicted by these empirical relations \citep{eva09}. One possible explanation for this discrepancy is that the dense gas most relevant to the star formation process appears diluted in extragalactic observations due to the more abundant diffuse clouds.  In contrast, the spatial resolution in Galactic star forming regions is sufficient for the the dense gas to fill the beam \citep{hei10}. 

Another important difference between the nearby, well-studied star forming regions and star forming regions observed in other galaxies is that the nearby regions are primarily forming low mass stars. Due to the feedback between massive stars and their environments, it is important to also study massive Galactic regions analogous to the star forming regions seen in other galaxies that are near enough to obtain an accurate census of star formation activity from high mass stars to below 1 M$_{\odot}$.

\begin{figure}
\includegraphics[width=0.9\textwidth]{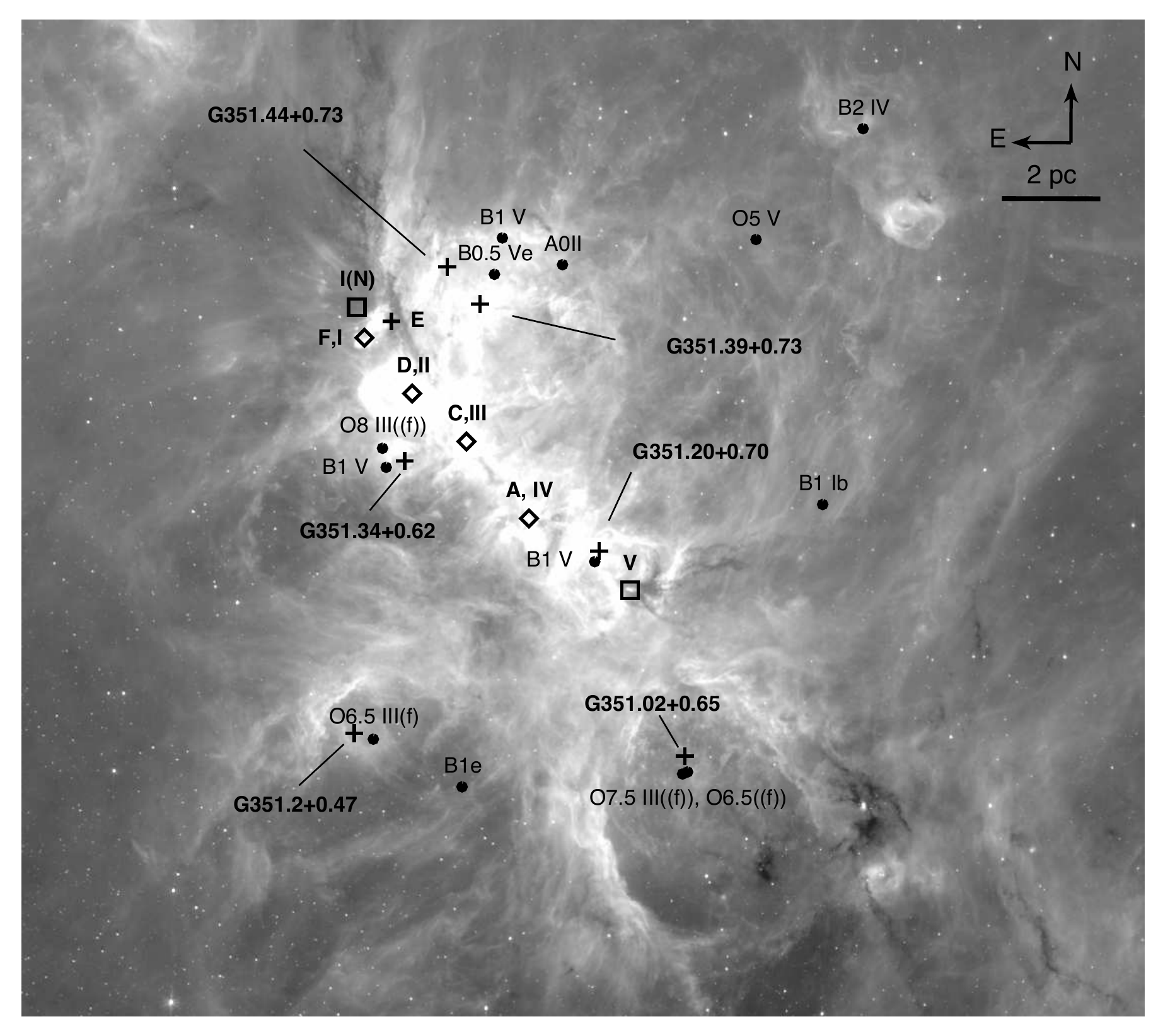}
\caption{\emph{Spitzer} IRAC 8.0 $\mu$m image of NGC 6334 with the FIR sources from \citet{mcb79} and \citet{gez82} (I, I(N), II, III, IV, V) and the radio sources (A, C, D, E, F) from \citet{rod82} marked. Sources identified only in the radio are marked with plusses, sources identified only in the far-infrared are marked with boxes, and sources that are detected in both the radio and FIR are marked with diamonds. The massive stars believed responsible for the excitation of the optical nebulae are also marked with dots and labeled by spectral type as given in \citet{per08} and references therein. The image is centered at 17$^h$20$^m$01$^s$ -35$^d$53$^m$11$^s$. The scale bar is approximately 4.2$'$ in length, or 2 pc at the assumed distance of 1.6 kpc.}\label{fig:NGC6334LABEL}
\end{figure}

One such Galactic region is NGC 6334, or the Cat's Paw Nebula. The current understanding of NGC 6334 is well-summarized in a recent review by \citet{per08}. NGC 6334 is a large (M $>$ 10$^5$ M$_\sun{}$) molecular cloud complex located at $l=351\degr{}, b=0.7\degr{}$. OB stars have been detected both in clusters and isolated throughout the entire molecular cloud complex, with a recent UBVR survey turning up 150 OB stars \citep{rus12}. The central portion of the cloud consists of a dense dusty ridge extending approximately 10 pc parallel to the Galactic Plane hosting multiple sites of massive star formation. Historically different nomenclature has been used to describe the brightest radio and far-infrared sources in this region. We are using the A, C, D, E, and F radio source labels from \citet{rod82}, the roman numberal I, II, III, IV and V FIR source names from \citet{mcb79} and the I(N) reference from \citet{gez82}. Sources that are detected in both the radio and far-infrared continuum we will refer to by both designations, e.g. NGC 6334 I/F. The main features of NGC 6334 are shown in Figure \ref{fig:NGC6334LABEL}. NGC 6334 I(N) and I/F are among the farthest north in the ridge and are also extremely young. I(N), a deeply embedded protocluster of massive stars, has recently been studied extensively with observatories such as Herschel and the Sub-Millimeter Array (SMA)\citep{bro09}. Just to the south is NGC 6334 II/D, a small well-defined HII region. NGC 6334 V is another young far-infrared source which is driving a massive molecular outflow \citep{fis82}. NGC 6334 III/C and NGC 6334 IV/A are also joint far-infrared and radio sources, each surrounding a late O-type star and having $L_{bol}>10^{5}L_{\sun}$. 

This diversity of environments within NGC 6334 provides a unique laboratory to study massive star formation at all stages of pre-main sequence evolution. In addition, as NGC 6334 is one of the closest sites of massive star formation to the Sun ($1.61\pm0.08$ kpc, \citealt{per08}), we are also able to characterize the large population of low mass stars the cloud is producing. Prior to this work, there have been no deep, extensive near- and mid- infrared surveys covering the entire NGC 6334 star forming complex in a systematic way. Shallow observations in the mid-infrared include the Galactic Legacy Infrared Mid-Plane Survey Extraordinaire (\emph{GLIMPSE}, \citealt{ben03,chu09}) and the MIPS Inner Galactic Plane Survey (MIPSGAL, \citealt{car09}) programs using the Infrared Array Camera \citep{faz04} and the Multiband Imaging Photometer for Spitzer (MIPS, \citealt{rie04}) onboard the Spitzer Space Telescope \citep{wer04}. Deep ground-based near-infrared observations have targeted small portions of the cloud, e.g. \citealt{tap96}, and shallow low resolution studies of the whole cloud have also been completed (e.g. \citealt{str89}). In this paper we present the analysis of new deep observations with Spitzer IRAC as well as a large deep survey of the entire star forming region in the near infrared with the National Optical Astronomy Observatory (NOAO) Extremely WideField InfraRed IMager (NEWFIRM) camera \citep{NFreference} on the Blanco 4m telescope at CTIO. We also incorporate our analysis of the publicly available 24 $\mu$m data from the MIPSGAL survey program.

In this comprehensive study we have obtained a census of the young stellar population of this massive and complex region extending from the high mass protostars to below $1 M_{\sun}$. We identify the sources based on the morphology of their spectral energy distribution (SED). Sources that have a rising spectrum toward long wavelengths are identified as Class I Young Stellar Objects (YSOs) and sources that have a flatter, or negative slope toward longer wavelengths are identified as Class II YSOs.

Combined with our precise mapping of extinction in the region, we have extended the empirical relations between gas surface density and star formation rate to a broader range of masses. This is the first paper of a series in which we will examine star formation efficiency in 5 other massive Galactic star forming regions that we have also observed with Spitzer and NEWFIRM.

In Section~\ref{sec:OBS} of this paper, we describe our observations and in Section~\ref{sec:DAOPHOT} we present our methodology for the data reduction and photometry. In Section~\ref{sec:YSOID} we present our method to identify the candidate young stellar object (YSO) population, and in Section~\ref{sec:Discussion} we delve into the derived properties of the star formation activity in NGC 6334.

\section{OBSERVATIONS}\label{sec:OBS}

\begin{figure}
\includegraphics[width=0.9\textwidth]{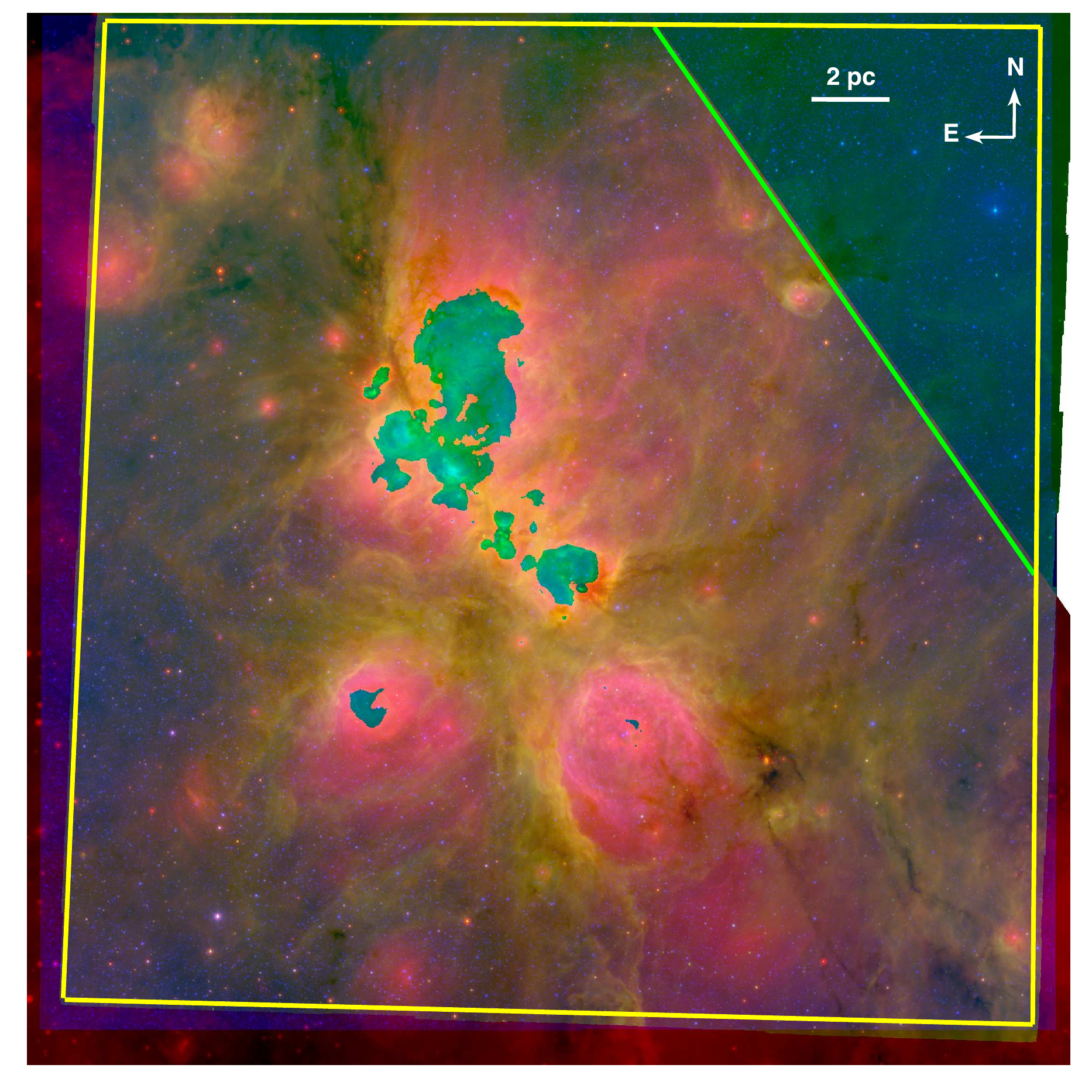}
\caption{Three color image of NGC 6334 with $J$ band for blue, 8.0 $\mu$m for green, and 24 $\mu$m red. The cyan colored regions near the center mark where the 24 $\mu$m mosaics are saturated. The region outlined in yellow denotes the area with full 7 band coverage, $J$, $H$, and $K_{s}$ band from NEWFIRM and 3.6, 4.5, 5.8 and 8.0 $\mu$m with IRAC. The area in the northwest corner above the green line is missing MIPS 24 $\mu$m coverage. The image is centered at 17$^h$20$^m$01$^s$ -35$^d$53$^m$11$^s$. The scale bar is approximately 4.2$'$ in length, or 2 pc at the assumed distance of 1.6 kpc.}\label{fig:ALLRGB}
\end{figure}

Figure~\ref{fig:ALLRGB} shows the footprint of our observations of NGC 6334. The cloud is observed in its entirety with our near-IR (NEWFIRM) and mid-IR (IRAC) data. Archival observations from the MIPSGAL survey cover all but the northwest portion of the cloud. Approximately 0.8 square degrees were observed from $J$ band to 8.0 $\mu$m.

\subsection{Spitzer Observations}

\begin{figure}
\includegraphics[width=0.9\textwidth]{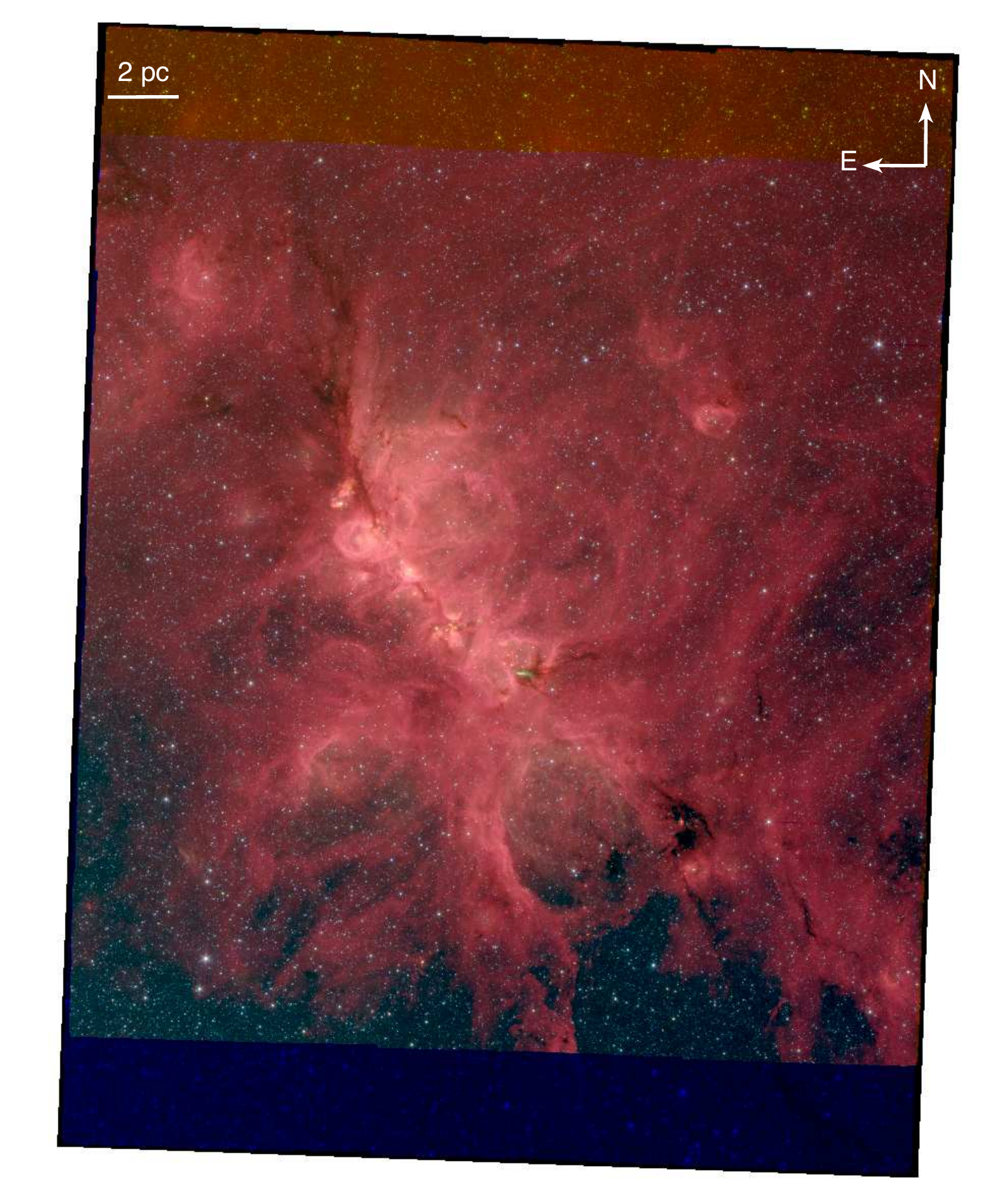}
\caption{Three color image of the observed IRAC field using the high dynamic range (HDR) mosaics in different bands. We have generated the HDR mosaics by smoothly merging the short frame mosaics to fill in the saturated areas in the long frame mosaics. The color coding is blue, green, red for $\lambda$= 3.6, 4.5, 8.0 $\mu$m respectively. The image is centered at 17$^h$20$^m$01$^s$ -35$^d$53$^m$11$^s$. The scale bar is approximately 4.2$'$ in length, or 2 pc at the assumed distance of 1.6 kpc.}\label{fig:IRACRGB}
\end{figure}

We observed a 1.2\degr{} by 0.9\degr{} region covering NGC 6334 (shown in Figure~\ref{fig:IRACRGB}) using IRAC during Spitzer's cryogenic mission on 2008 September 21 (PID 30154) at 3.6, 4.5, 5.8, and 8.0 $\mu$m. We mapped the region with a grid of 13 by 11 tiles in 2 epochs spaced by 6 hours with 3 dither positions per tile per epoch. The two epochs were intented to be scheduled 6 months apart (to allow variability studies) but were then executed back to back to ensure completion of the dataset before exhaustion of the cryogen. The region observed corresponds to approximately 33 by 25 parsecs at a distance of 1.6 kpc. 

We used IRAC's 12 s frame time High Dynamic Range (HDR) mode, acquiring consecutive individual observations with exposure times of 0.4 and 10.4 seconds. The long exposures probe the faint stellar population in the cloud while the addition of the short exposure observations allow for the recovery of bright sources that are saturated in the long exposure frames. Due to the short interval between the two epochs, both epochs were combined to produce the final mosaics. The 858 individual Basically Calibrated Frames (BCD, processed with the Spitzer IRAC pipeline version S18.7.0) for each exposure time were mosaicked together using the IRACproc package \citep{sch06} to perform outlier rejection and bad pixel masking. IRACproc is a PDL wrapper script for the standard Spitzer Science Center mosaicing software MOPEX \citep{MOPEXCITATION} which has been enhanced for better cosmic ray rejection. Each mosaic was re-sampled from the native IRAC resolution of 1.22\arcsec{} per pixel to 0.86267\arcsec{} per pixel and projected onto a common world coordinate system grid. The final mosaics combined a median of 6 dither positions per pixel for an effective integration time of 62.4 seconds per pixel in each long exposure mosaic and 2.4 seconds in the short exposure mosaic. The bright nebular emission along the central ridge of NGC 6334 severely saturated the central portion of the long exposure mosaic at 8.0 $\mu$m but remains unsaturated in the short exposure mosaic.

We also retrieved the archival MIPSGAL 24 $\mu$m observations of NGC 6334 obtained as part of the MIPSGAL survey.  The central ridge running from the northeast to the southwest of NGC 6334 is saturated in the MIPSGAL 24 $\mu$m images. The available observations of NGC6334 span across two MIPSGAL mosaics, shown combined in Figure~\ref{fig:ALLRGB}. The upper northwest corner of NGC 6334 ($b>1.0$) was not covered by the MIPSGAL survey. Full details of the MIPSGAL observations can be found in \citet{car09}. The PSF of the 24 $\mu$m mosaics is~6\arcsec{} FWHM, and the median exposure per pixel in these MIPSGAL mosaics was 30 s.
 
\subsection{NEWFIRM Observations}

\begin{figure}
\includegraphics[width=0.9\textwidth]{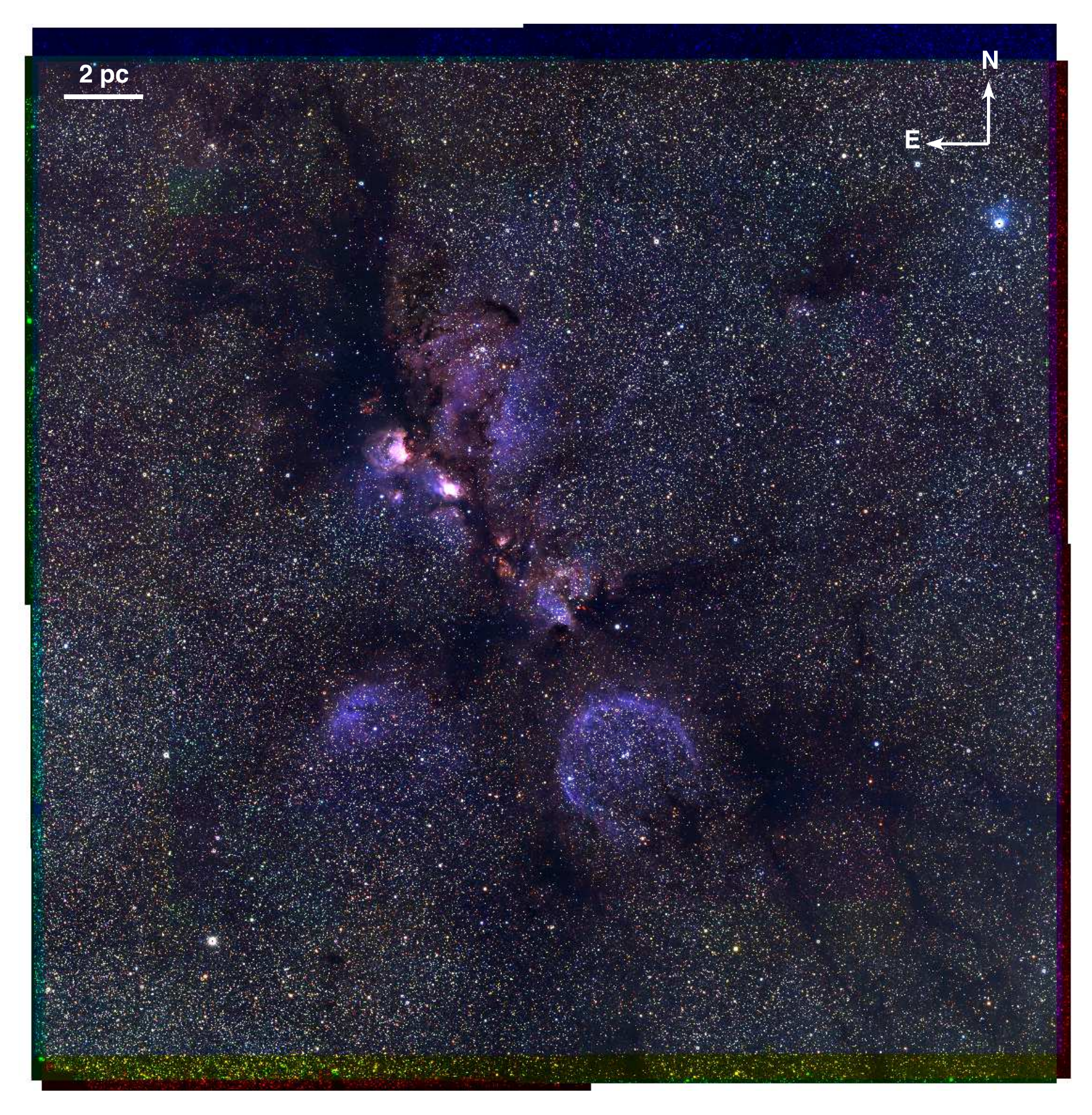}
\caption{Three color image of the central pointing NEWFIRM field. The color coding is blue, green, red for J, H, and $K_{s}$ band respectively.  The image is centered at 17$^h$20$^m$01$^s$ -35$^d$53$^m$11$^s$. The scale bar is approximately 4.2$'$ in length, or 2 pc at the assumed distance of 1.6 kpc.}\label{fig:NEWFIRMRGB}
\end{figure}

Figure~\ref{fig:NEWFIRMRGB} shows our near-IR images of NGC 6334 acquired with the NEWFIRM camera on the Blanco 4 m telescope at Cerro Tololo Inter-American Observatory (CTIO) during the 2010A semester between 2010 May 24 and 2010 May 30. The NEWFIRM camera, designed to quickly map large areas of the sky, contains 4 InSb 2048 $\times$ 2048 pixel arrays arranged in a 2 $\times$ 2 pattern with a 28\arcmin{} field of view and an approximately 1\arcmin{} gap between the CCDs. The detector has a pixel scale of 0.4\arcsec{} per pixel.  

To match the 1.2\degr{} by 0.9\degr{} sky coverage of our IRAC mosaics, we used a 2 by 2 central pointing mapping routine to divide NGC 6334 into four quadrants plus an overlapping central region. Our observations were taken with a random dither offset large enough to fill in the 1\arcmin{} gap between CCDs. For each quadrant we took 60 second exposures at 11 dither positions. Each dither position was a combination of multiple co-added exposures to achieve longer exposure times while avoiding the non-linear regime of the detector. For the $J$ band we used two co-added 30 second exposures, for the $H$ band we used three co-added 20 second exposures, and for the $K_{s}$ band we used six co-added 10 second exposures. In the central pointing, the $J$ band image used 20 second exposures and 3 co-adds, and the $H$ and $K_{s}$ bands used a slightly shallower 4 second exposures and 8 co-adds. The sky background was measured by taking 3 dithered sky observations 2\degr{} off field. 

Standard processing of dark frame subtraction, flat fielding, sky-subtraction and bad pixel masking was performed by the NEWFIRM Science Pipeline \citep{dic09,swa11} to produce 5 stacked composite images for each near-infrared band. The final stacked images retain the detector pixel scale of 0.4\arcsec{} per pixel. Variable seeing conditions caused the final observed point source FWHM in the stacked images to range from approximately 0.9\arcsec{} to 1.5\arcsec{}. The $H$ band observations were taken during the best seeing conditions and have the smallest FWHM, and the $K_{s}$ images have the largest FWHM. Even in the most crowded portions of the region observed, however, the difference in effective resolution at the different NEWFIRM bands has a minimal effect on later source-matching.

\section{SOURCE FINDING AND PHOTOMETRY}\label{sec:DAOPHOT}

We used DAOPHOT \citep{ste87} to perform PSF-fitting photometry in each final mosaic individually; the long and short exposure mosaics for each of the 4 IRAC bands, the 5 pointings per band from NEWFIRM, and the 2 available MIPSGAL 24 $\mu$m mosaics. 

The PSF for each image was derived from a minimum of 10 bright, unsaturated, and relatively isolated stars that are well separated from the extended nebular emission. The PSF obtained for the short-exposure IRAC frames was also used for the corresponding long exposure mosaic. The instrumental PSF was defined in a 24.4\arcsec{} $\times$ 24.4\arcsec{} box (20$\times$20 IRAC native pixels) chosen to be approximately 10 times larger than the observed FWHM of an individual point source and therefore include the vast majority of the flux of a point source.

Due to the crowded Galactic field at NGC 6334's latitude ($b=0.7\degr{}$), even in the short exposure mosaics most of the bright sources are surrounded by other point sources within the 24.4\arcsec{} box used to define the PSF. If left alone, the flux from these nearby point sources will skew the derived PSF and introduce additional systematic error into the photometry. We utilized a multiple-iteration method to converge on a satisfactory PSF for each frame. We began with a PSF composed purely of an analytical Lorentz function and used it to subtract off the neighboring sources from each of the 10 or more selected PSF stars. The source-subtracted image was then re-examined, and the PSF re-derived, allowing corrections to the analytic function based on the observed shape of the PSF. We progressed through four total iterations of fitting and subtracting the neighboring sources and then redefining the PSF from the PSF stars in the neighbor-subtracted frames.

The high variability of the background nebular emission at IRAC wavelengths over small spatial scales proved challenging for a single point source detection threshold to be used across the entire mosaic. To better enable the detection of all faint point sources regardless of position in the mosaic, we used a multiple-iteration background subtraction method similar to the procedure in \citet{str10}. We first used the photometry for the preliminary source list to subtract all point sources from each frame. The source-subtracted image was then median filtered with a 51 $\times$ 51 pixel box to produce an image of the background emission. This background image was then subtracted from the original mosaic. The background subtracted mosaics were then used to refine the location of the point sources and identify the ones that were lost in the background diffuse emission. For our NEWFIRM images, the variability of the sky background was low enough that we did not need to use the background subtraction method to achieve comparable completeness. 

\begin{figure}
\includegraphics[width=0.9\textwidth]{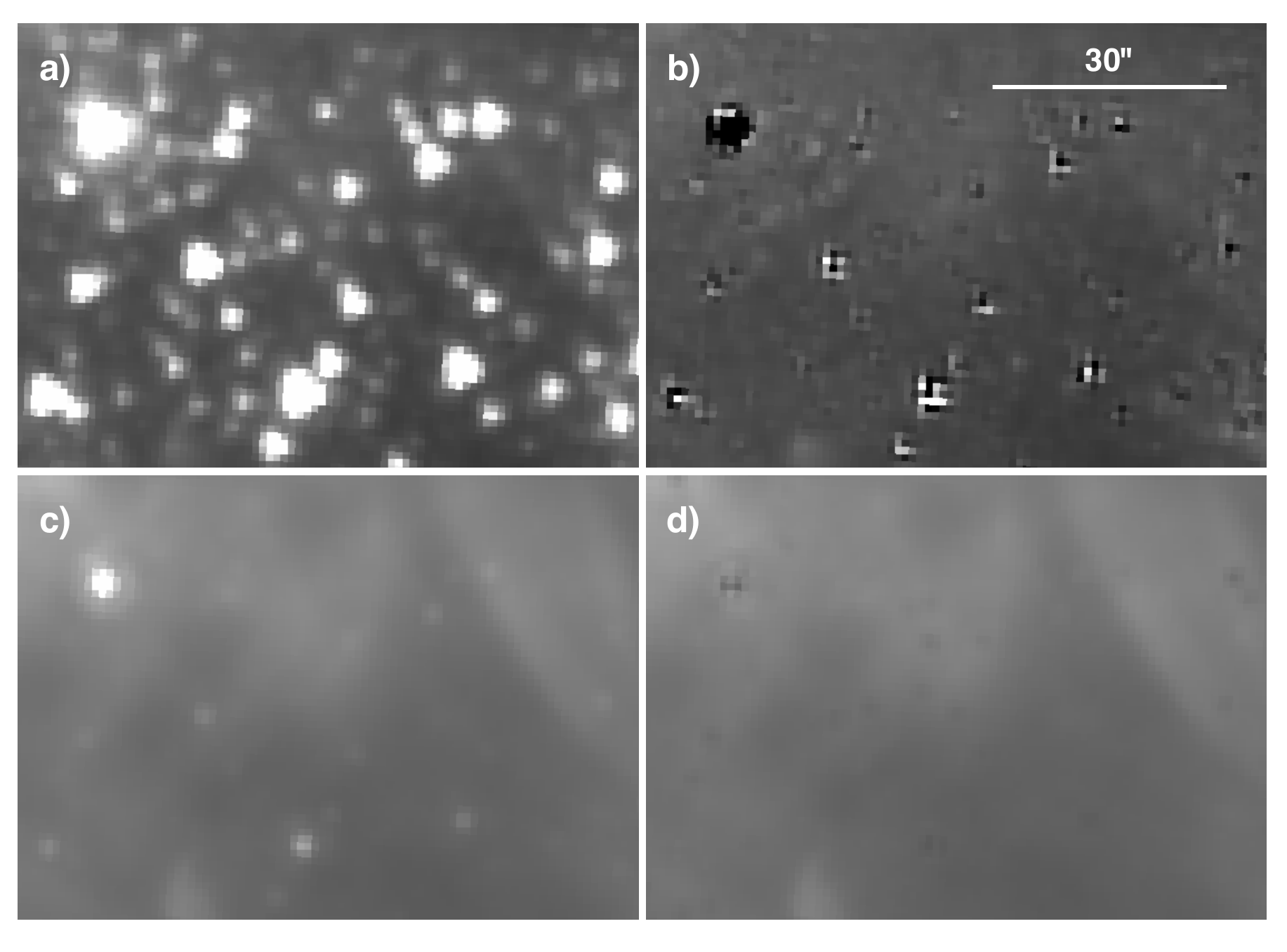}
\caption{An example of PSF subtraction in a representative crowded field in NGC 6334. The left panels show the long exposure mosaics at $\lambda$=3.6 $\mu$m (a) and 8.0 $\mu$m (c). The right panels show the result of point source subtraction using the PSF derived from the isolated bright stars in each corresponding short exposure mosaic for $\lambda$=3.6 $\mu$m (b) and 8.0 $\mu$m (d).}\label{fig:SourceSubtraction}
\end{figure}

The resulting source list was fed into DAOPHOT's \emph{ALLSTAR} routine to PSF-fit each source. To reduce the effect of crowding, only the flux within a 3 pixel radius was used to fit the model PSF, corresponding to 2.6\arcsec{} in the IRAC images and 1.2\arcsec{} in the NEWFIRM images. Detected sources which deviate from the expected point source shape by more than $5\%$ were rejected. The majority of rejected sources were spurious detections along the extended emission, although they also can indicate partially resolved background galaxies as well as blended PSFs from pairs of stars with extremely small angular separation. An example of the residuals left after subtracting the derived PSF for a representative crowded field can be seen in Figure~\ref{fig:SourceSubtraction}.

We ran the subtracted mosaics through DAOPHOT for two subsequent passes to find point sources that were missed during the first pass. Occasionally when running a second or third iteration, the DAOPHOT \emph{find} algorithm would detect bright pixels in the residuals of subtracted sources as a new legitimate point source. In all bands, any source detected in the second or third photometry iteration that had a centroid position matching within less than 2\arcsec{} (approximately the mean FWHM of the IRAC point sources) of a source from a previous iteration was considered a duplicate detection. The duplicate sources identified in this way were rejected from further analysis.

Sources brighter than magnitude 9.2 are saturated in the long exposure mosaic at 3.6 $\mu$m. The saturation limit is 9.0, 6.5 and 6.5 magnitudes for 4.5, 5.8 and 8.0 $\mu$m respectively. Where a source in the long exposure mosaic frame matched within 2.0\arcsec{} of a source with a short exposure magnitude brighter than the saturation limit, the short exposure photometry was used. 
 
The same method described above was used to compile a preliminary source list for each of the five images for the three near-infrared bands, $J$, $H$, and $K_{s}$. In each photometric band, the NEWFIRM pointings overlapped by approximately 3\arcmin{}, causing many sources to be detected in more than one NEWFIRM stacked image. Sources with centroid positions matching within 0.8\arcsec{} in different stacked images were combined; we retained the average magnitude weighted by the exposure time contributing to that source in each stacked image. Point sources detected in 2 or fewer stacked images (out of the maximum possible 11 for each mosaic) were rejected as being likely artifacts left over after pipeline processing. As a result, 120 seconds is the minimum exposure time for each NEWFIRM source included in the point source catalog. The median point source exposure time was 600 seconds for $J$ and $K_{s}$ band and 480 seconds for $H$ band.

We also applied the same photometry process to the MIPSGAL 24 $\mu$m mosaics. Similar to the 8.0 $\mu$m mosaics from IRAC, the \emph{find} algorithm identified many small bright areas along the extended emission as point sources. Larger restrictions were placed on the DAOPHOT PSF-fitting parameters of detected 24 $\mu$m point sources to decrease the number of spurious detections. After these cuts 991 point sources in the MIPSGAL mosaics have a centroid matching within 1\arcsec{} of an IRAC/NEWFIRM source. Due to the issues with saturation, there is a large difference in the completeness of the MIPS portion of the source catalog across the region.

We used the short exposure mosaic to calibrate the instrumental PSF photometry in each IRAC band against 100 relatively isolated and bright stars measured with aperture photometry in PhotVis \citep{gut04}. The same procedure was used to calibrate the 24 $\mu$m photometry. To determine the zero point for our near-infrared photometry, we matched up bright isolated stars from our NEWFIRM catalogs with sources from the Two Micron All Sky Survey (2MASS, \citealt{skr06}) source list. In the higher source-density portions of the field the 2MASS photometry suffers more from crowding and blending due to the lower survey resolution (2.4\arcsec vs. 0.9\arcsec), leading to a larger number of sources with artificially bright magnitudes. To decrease the flux contamination effect from this crowding, sources for the near-infrared photometric calibration were carefully chosen from uncrowded portions of the mosaics. 

\subsection{NGC 6334 Point Source Catalog}\label{sec:PSC}
We created the final source list catalog by spatially matching the detected sources across the different bands. We first matched sources detected with the same instrument, $J$, $H$ and $K_{s}$ for NEWFIRM and 3.6, 4.5, 5.8, and 8.0 $\mu$m for IRAC. Our cross-band correlation method first calculates whether there is an overall systematic shift in source centroid positions between the input frames, and then looks for sources with shifted centroid positions matching between bands. The NEWFIRM sources were matched if their centroid positions differed by less than 0.8\arcsec{}. For IRAC, due to the larger pixel scale and broader FWHM, we required sources to have centroids that matched within 1.0\arcsec{} between bands. To combine the two catalogs together, we used the larger 1.0\arcsec{} matching criterion.

\begin{figure}
\includegraphics[width=0.9\textwidth]{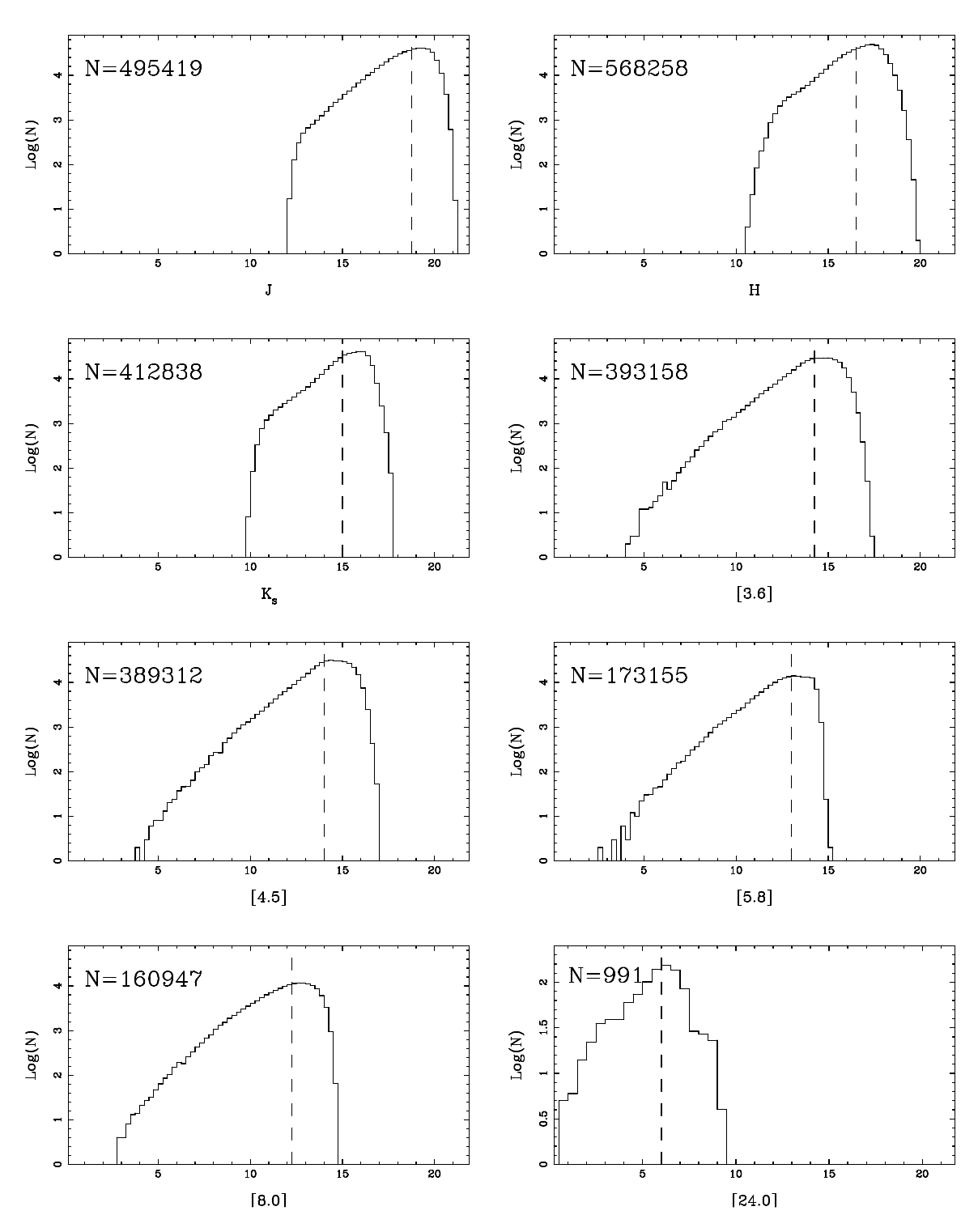}
\caption{Source Histograms for bands $J$ through 24 $\mu$m showing the limiting magnitude and completeness limit for each band.}\label{fig:SourceHist1}
\end{figure}

To further increase reliability of our point source catalog, we first required simultaneous detection in 2 adjacent bands (e.g. $H$ and $K_{s}$ on NEWFIRM, or 3.6 and 4.5 $\mu$m with IRAC). Following that, we also included sources at the edges of the IRAC mosaics only visible at 3.6 and 5.8 $\mu$m or at 4.5 and 8.0 $\mu$m bands. Finally, due to the large number of very red sources in the field, we also retained photometry for approximately 7,000 $K_{s}$ band sources lacking detection at $J$ or $H$ but that matched within 0.5\arcsec{} with a source detected in 2 or more adjacent IRAC bands. Histograms of all sources kept for the NGC 6334 point source catalog are presented in Figure~\ref{fig:SourceHist1}. The completeness limit for each band was estimated from the turnover point in each histogram and the limiting magnitude reported is the faintest source detected at 5$\sigma$ level or higher (a maximum photometric error of 0.217 magnitudes). Sample entries of the NGC 6334 Point Source Catalog can be seen in Table~\ref{tab:PSCSAMP}. The full version is available in electronic format. A statistical summary of the Point Source Catalog can be found in Table~\ref{tab:CatSumFull}.

\section{YSO IDENTIFICATION}\label{sec:YSOID}

Observational methods of identifying YSOs have been derived from evolutionary models and comparison to observed known YSOs. One early method was to fit the slope $\alpha$ of log($\lambda$F$_{\lambda}$) vs. log($\lambda$) between 2 and 24 $\mu$m \citep{lad87}. Other methods have referenced color selection criteria for YSO stars, (e.g. \citealt{gut09}). We used the full range of observed wavelengths from $J$ band to 24 $\mu$m for the purpose of identifying YSO candidates in our point source catalog. The selection is based on color and magnitude criteria, similar to those defined in \citet{gut09}. We have however adapted these selection criteria to take into account the greater distance of NGC 6334 and the higher level of contamination due to crowding. Furthermore, we have developed and validated new criteria designed to allow selection using only near-IR and warm Spitzer bands (3.6 and 4.5 $\mu$m). These new criteria are important for our follow-up programs on other high mass star-forming regions that have been observed after Spitzer's liquid helium cryogen supply was exhausted as well as for other massive regions where existing 5.8 and 8.0 $\mu$m IRAC observations are compromised by saturation from the very strong PAH extended emission.

To identify the YSOs in NGC 6334 we have first applied the original \citet{gut09} criteria to sources with detection at all four IRAC bands, then applied our newly developed criteria to the combined near- and mid-IR point source catalog. As a comparison, we then compare the results of these YSO identification methods to the classification devised by fitting a power law to the SED between 2 and 24 $\mu$m.

\subsection{IRAC 4 Band Detected Sources}\label{sec:4IYSO}
The first stage of our YSO identification followed the procedure outlined in \citet{gut09}. This method looks at a variety of color-color and color-magnitude relations for sources detected in all 4 IRAC bands. First, this method identifies non-YSOs which share some similar color spaces to YSOs and that could contaminate the final source statistics if not removed. These possible contaminants include active galactic nuclei, quasars, resolved knots of shocked gas emission, and faint stellar point sources where the longer wavelength photometry is contaminated by PAH emission, which is a particularly common cause of contamination in this massive star forming region. 

\begin{figure}
\includegraphics[width=0.9\textwidth]{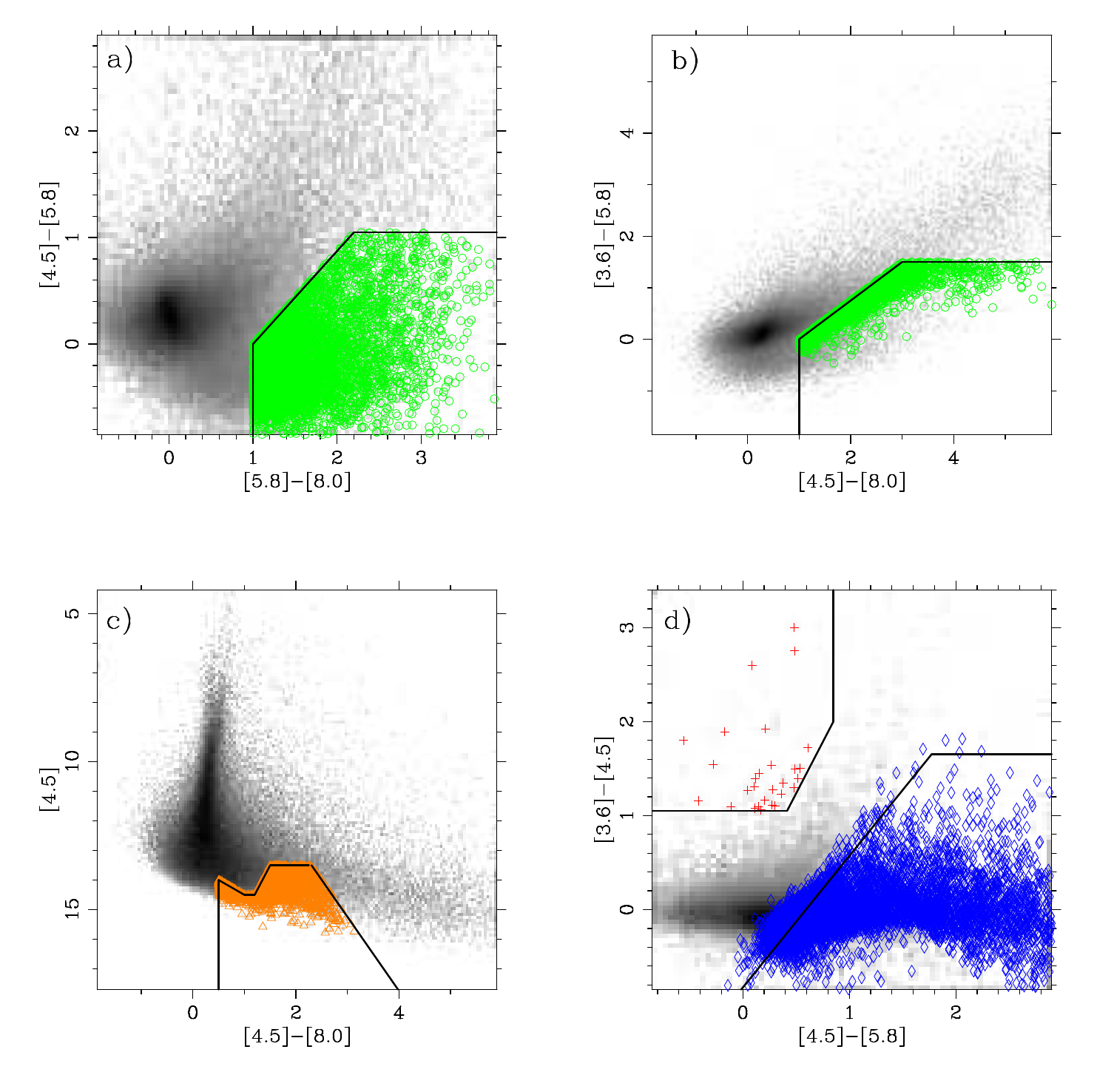}
\caption{Color-color diagrams used to identify contaminant objects among the sources with detection at all 4 IRAC bands following the criteria in \citep{gut09}. The background logarithmic gray-scale indicates the overall source density in each color-color and color-magnitude space. In panels a) and b) PAH galaxies are marked with green circles. In panel c) candidate AGNs are marked by orange orange triangles. Panel d) shows knots of shocked emission (red plusses) and PAH contaminated sources (blue diamonds). A color version of this figure is available online.\label{fig:IRAC4BANDCONT}}
\end{figure}

\begin{figure}
\includegraphics[width=0.9\textwidth]{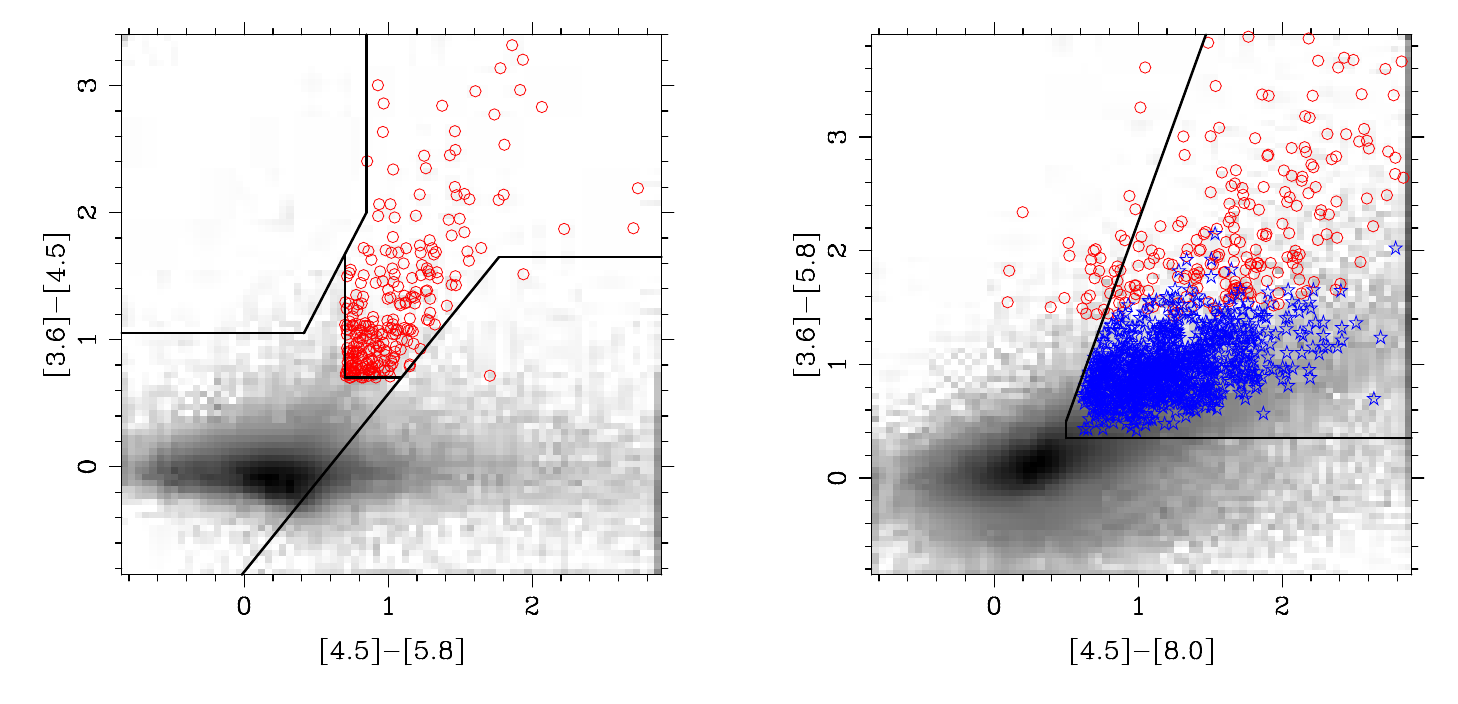}
\caption{Class I (red circles) and Class II (blue stars) YSOs identified using the \citet{gut09} selection criteria for sources with detection in all 4 IRAC bands. The background logarithmic gray-scale image shows the distribution of all sources detected at all 4 IRAC bands. The left panel is used for the identification of Class I YSOs. Sources in the region outlined in the right panel not previously identified as Class I YSOs or contaminant objects are identified as Class II YSOs. A color version of this figure is available online.\label{fig:IRAC4BANDYSO}}
\end{figure}

We applied this classification scheme to the approximately 80,000 point sources in our catalog with 5$\sigma$ detection at all 4 IRAC bands. We found that 10,327 sources were likely background galaxies, 41 sources were likely knots of shocked gas emission, and 13,791 sources were identified as faint sources with PAH contaminated apertures at long wavelengths. After the decontamination process we were left with 260 sources matching the color selection criteria for Class I YSOs, and 915 sources were identified as Class II YSOs. The application of each step of the full Gutermuth contaminant object identification criteria is shown in Figure~\ref{fig:IRAC4BANDCONT} and the identified YSOs are shown in Figure~\ref{fig:IRAC4BANDYSO}.

\subsection{Near-Infrared and 3.6 and 4.5 $\mu$m Sources}\label{sec:NIRYSO}

The SED of a YSO is characterized by excess emission above expected photospheric levels in the infrared, increasing towards longer wavelengths through the near- and mid-infrared. However, in regions where massive star formation is ongoing, the extreme levels of ionizing radiation from the massive O and B stars excites widespread emission from PAH compounds embedded within the dust clouds. This extended emission dominates the observed emission at IRAC's 5.8 and 8.0 $\mu$m bands, drowning out the individual point sources nestled within these structures and saturating the detector in even modestly long integration times. For our second stage of our YSO selection process we identified additional YSO sources that lacked detection at 5.8 or 8.0 microns and were thus missed by the \citet{gut09} criteria. We used the combined NEWFIRM and IRAC near-infrared catalogs for the over 260,000 sources detected at $H$, $K_{s}$, 3.6 $\mu$m and 4.5 $\mu$m. As mentioned before, these same criteria are also suitable for YSO selection in fields observed during the Spitzer warm mission. 

For this stage we first identified the locus of main sequence stars (assuming an average spectral type of K0) in the $K_{s} - [3.6]$ vs. $[3.6] - [4.5]$ and $H - K_{s}$ vs. $K_{s} - [4.5]$ color spaces subjected to interstellar reddening. We adopted the parameter $\sigma_{c}$, defined as the maximum error in color for two sources each having the maximum catalog cut-off photometric error of 0.217 magnitudes ($S/N\sim5$), or $\sigma_{c}=0.307$. Sources falling within $\sigma_{c}$ of the main sequence reddening vector were considered to be reddened field stars.

\begin{figure}
\includegraphics[width=0.9\textwidth]{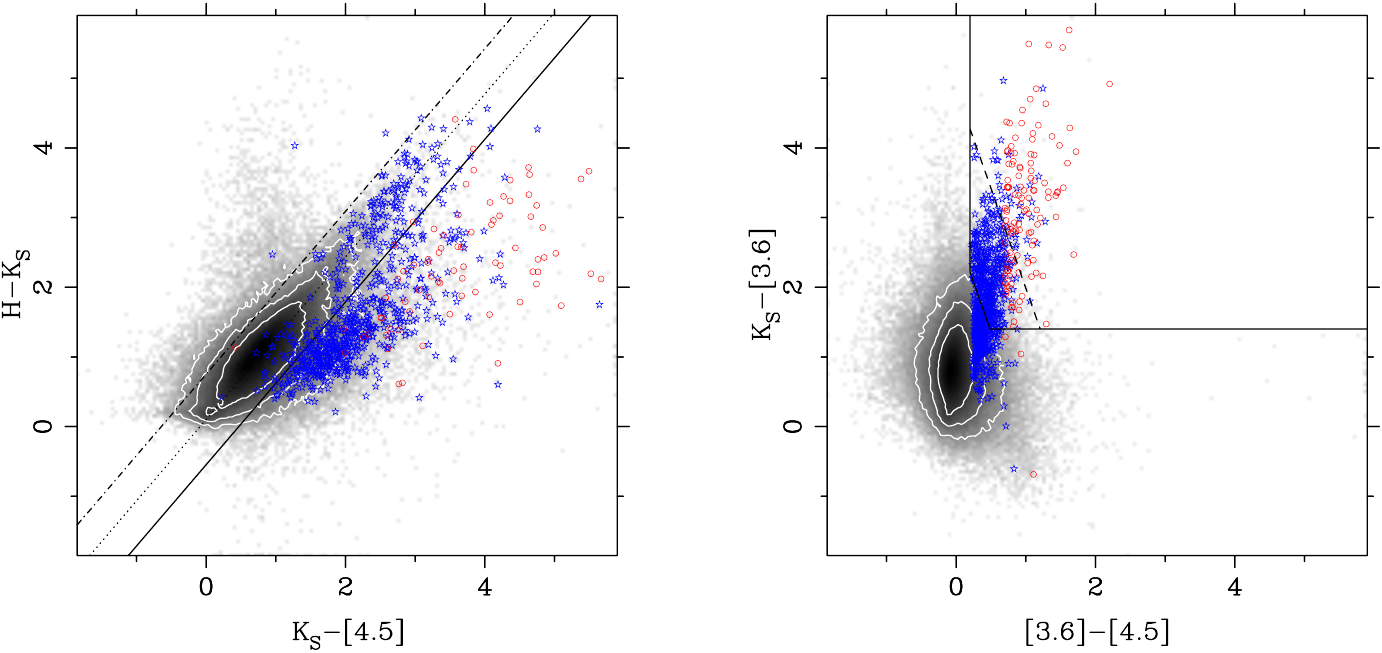}
\caption{The near-infrared colors spaces used to identify additional YSO candidates that were not detected in all 4 IRAC bands. In both panels, the background shows the distribution of all sources detected at $H,K_{s},[3.6],[4.5]$, and the white contours demarcate the 36, 86 and 95\% source density levels.  The red squares mark IRAC-selected Class I YSOs and the blue squares mark IRAC-selected Class II YSOs. On the left, the dotted line indicates the reddened main sequence locus. The dot-dash and the solid line indicate 1 $\sigma_{c}$ above and below the reddened main sequence, respectively. Sources more than 1 $\sigma_{c}$ below the main sequence are potential YSO candidates. On the right, the black lines designate the color space corresponding to the IRAC YSOs beyond the reddened main sequence in the upper right quadrant. The dark dashed line in the right hand panel marks the cutoff we have adopted in the near-infrared for Class I vs. Class II YSOs. A color version of this figure is available online.}\label{fig:BULLSEYE}
\end{figure}

We also considered where in these color spaces the sources already identified as YSOs based on the \citet{gut09} 4 IRAC band classification would fall, and the near-infrared $K_{s}-[3.6]$ vs. $[3.6]-[4.5]$ criteria also presented in \citet{gut09}. We identify the region beyond the identified 3$\sigma_{c}$ reddened main sequence star locus and bounding the IRAC-selected YSOs as containing the YSO candidate stars. The equations below describe the bounding regions containing YSO candidate stars and are shown in Figure~\ref{fig:BULLSEYE}.

\begin{equation} [3.6]-[4.5]>0.2 \end{equation}
\begin{equation} K_{s}-[3.6]>1.4 \end{equation}
\begin{equation} K_{s}-[3.6]>-2.857([3.6]-[4.5]-0.101)+2.5 \end{equation}
\begin{equation} H - K_{s} < 1.168(K_{s} - [4.5])-0.526 \end{equation}

\begin{figure}
\includegraphics[width=0.9\textwidth]{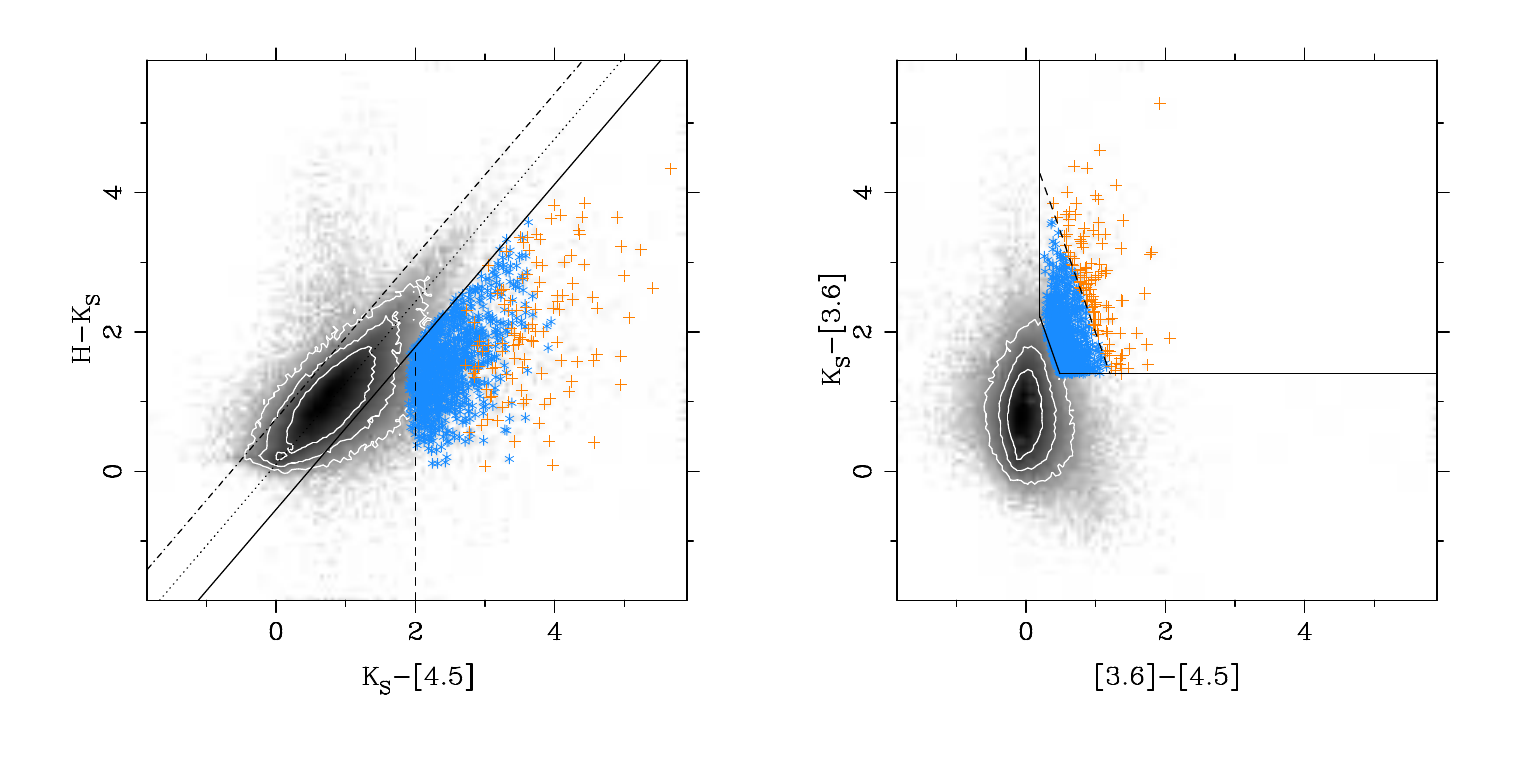}
\caption{YSO candidates that are identified by their near-infrared colors. Gray-scale and contours are the same as in Figure~\ref{fig:BULLSEYE}. Newly identified YSOs are marked as plusses, Class I YSOs in orange and Class II YSOs in cyan. A color version of this figure is available online.}\label{fig:NIRYSOCUTS}
\end{figure}

Sources that matched all of the above criteria are considered YSO candidates and are shown in Figure~\ref{fig:NIRYSOCUTS}. Sources with extremely red colors matching the constraints below are further identified as candidate Class I YSOs, while the rest that only meet the first set of near-infrared criteria are identified as Class II YSO candidates.

\begin{equation} K_{s}-[3.6]>-2.857([3.6]-[4.5]-0.401)+3.7 \end{equation}
\begin{equation} K_{s}-[4.5]>2 \end{equation}

\begin{figure}
\includegraphics[angle=0,width=0.9\textwidth]{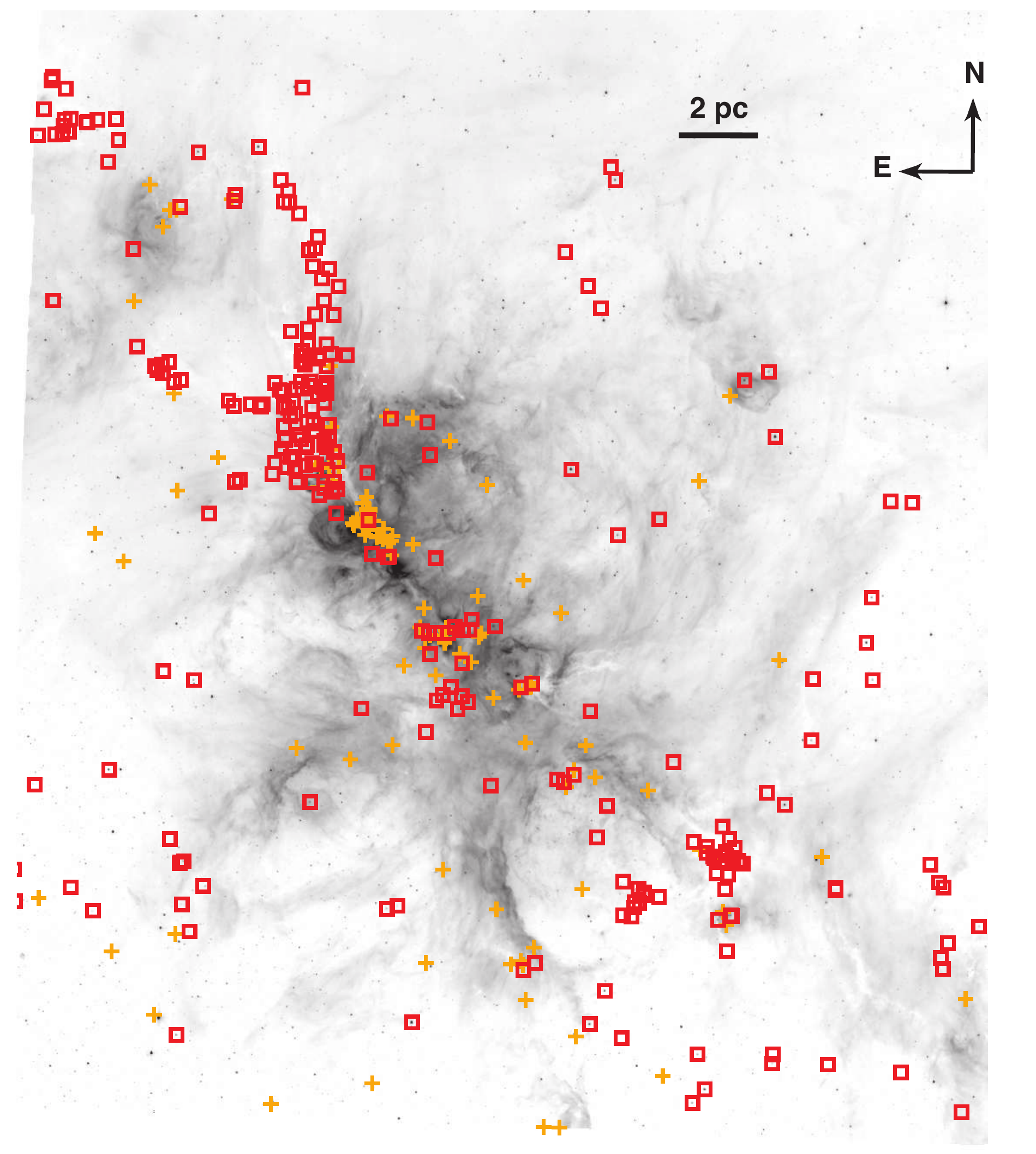}
\caption{The spatial distribution of all the Class I YSO candidates plotted over the IRAC 8.0 $\mu$m emission. The IRAC-selected candidates are plotted as red squares and the candidates added using the near-infrared criteria are shown as orange plusses. Many Class I sources in the bright nebulous regions of NGC 6334 that saturated at long IRAC observations are recovered by the near-infrared criteria. A color version of this figure is available online.}\label{fig:CLASSIYSOMAP}
\end{figure}

\begin{figure}
\includegraphics[angle=0,width=0.9\textwidth]{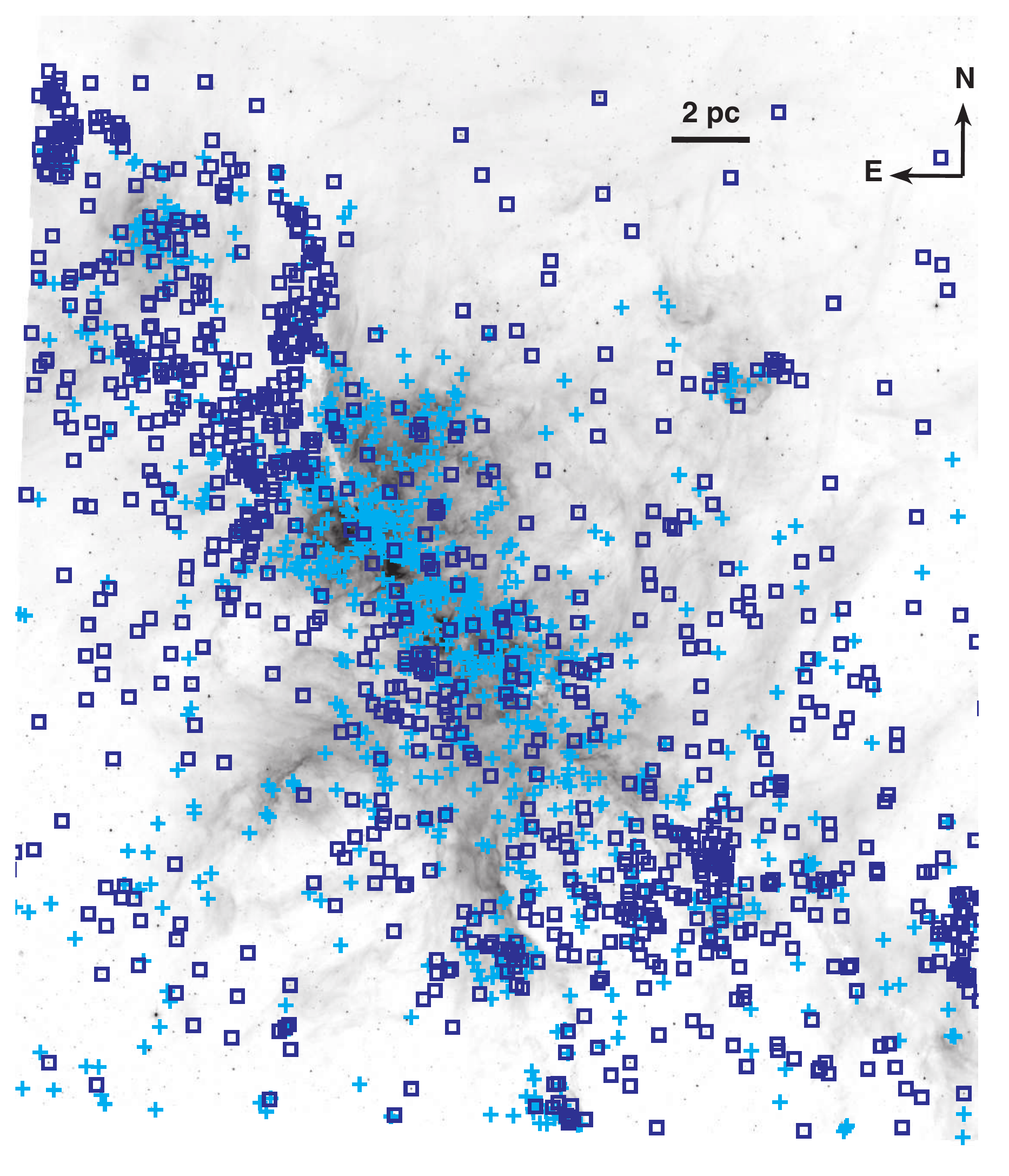}
\caption{The spatial distribution of all the Class II YSO candidates plotted over the IRAC 8.0 $\mu$m emission. The IRAC-selected candidates are plotted as blue squares and the candidates added using the near-infrared criteria are shown as cyan plusses. The additional Class II sources identified by the near-infrared criteria display show a more strongly clustered distribution than the IRAC 4 band sources, a consequence of the stricter criteria employed for near-infrared YSO selection. A color version of this figure is available online.}\label{fig:CLASSIIYSOMAP}
\end{figure}

Using these criteria, \nci{} Class I candidates and \ncii{} Class II candidate YSOs are added to the previously identified (4 IRAC bands) YSOs. The spatial distribution of the YSO candidate sources throughout the field can be seen in Figures~\ref{fig:CLASSIYSOMAP} and~\ref{fig:CLASSIIYSOMAP}. Viewed in projection against the cloud, the location of Class I sources is strongly correlated to the high density regions which show up both as the infrared dark clouds (IRDCs) and the bright filaments in the 8 $\mu$m IRAC image. The Class II sources are also strongly clustered, although a larger number of them are found further from the filamentary structure of NGC 6334. Although some additional spread in Class II location may be expected due to their expected more advanced age, this may also be an indication of the higher contamination fraction present in the Class II population.

Overall, the near-infrared cuts are somewhat more restrictive than the selection requiring 4 IRAC bands. This decreases the amount of contamination from sources extincted only by interstellar and not circumstellar material. In Figure~\ref{fig:BULLSEYE} a significant number of IRAC-identified Class II YSOs can be seen within the color spaces belonging to reddened field stars.  In total, 81 (31\%) IRAC-selected Class I YSOs and 707 (78\%) IRAC-selected Class II YSOs are also detected at both $H$ and $K_{s}$ band. Within this sample, 6\% of Class I YSO candidates do not meet the adopted near-infrared criteria and 62\% of the IRAC-selected Class II YSO candidates fail at least one of the near-infrared YSO criteria. These sources are within 3$\sigma$ of the reddened main sequence locus and many are co-aligned with high extinction features seen in the derived NGC 6334 extinction map (see Section~\ref{sec:ExtMap}). These sources have a high probability to be reddened sources, either Class III YSOs / pre-main-sequence stars associated with NGC 6334 or unrelated background Galactic stars. We have not attempted a detailed characterization of the Class III / pre-main sequence population in this work due to the difficulty inherent in distinguishing a small physical infrared excess from the scatter due to photometric error and matching between different beam sizes in this crowded field.

Due to the slight differences in the near-infrared and IRAC 4 bands YSO identification methods in the overlapping $3.6 - 4.5$ $\mu$m region, there is some mismatching between classification schemes. Of the 81 IRAC-selected Class I YSOs that were detected at both $H$ and $K_{s}$, 24 (30\%) are identified by the near-infrared criteria as Class II sources. Of the 707 IRAC-selected Class II YSOs detected at $H$ and $K_{s}$, 18 (2.5\%) are identified by the near-infrared criteria as Class I sources. For consistency, we retain the IRAC 4 bands classification for all of these sources.  

\begin{figure}
\includegraphics[width=0.9\textwidth]{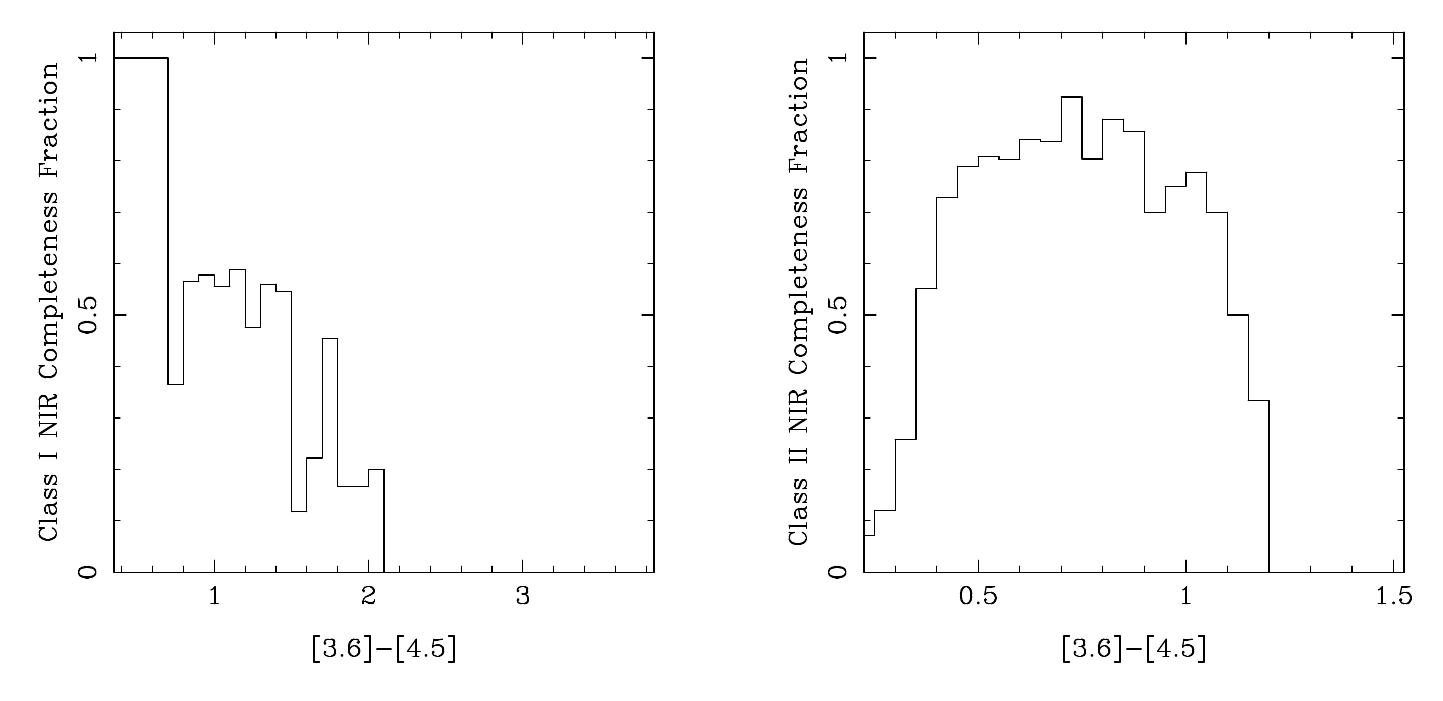}
\caption{The fraction of IRAC-selected Class I YSOs (left) and Class II YSOs (right) that are identified by only the near-infrared identification criteria.} The reddest sources are often lacking detection at either $H$ band or $K_{s}$ band and are therefore missed by the near-infrared identification criteria. The drop off of the Class II YSOs at $[3.6]-[4.5]<0.5$ demonstrates that the near-infrared YSO identification criteria are more restrictive in removing sources that may be reddened photospheres or Class III YSOs.\label{fig:NIRCOMPHIST}
\end{figure}

Combined, our near- and mid-infrared color criteria identified 375 Class I YSOs and 1908 Class II YSOs associated with NGC 6334. The completeness of the near-infrared criteria depends on the color of the source, with the fraction of YSOs recovered decreasing towards redder colors. Figure~\ref{fig:NIRCOMPHIST} shows the distribution in $[3.6]-[4.5]$ for the sources that are classified by both the IRAC 4-band and near-infrared classification schemes as the same YSO class. It is apparent that the near-infrared classification misses out on the reddest sources which are too deeply embedded to be detected at $H$ or even $K_{s}$. The downturn of the Class II source distribution below $[3.6]-[4.5]<0.5$ also suggests that the near-infrared criteria are successfully selecting against lightly reddened sources where the majority of contaminants would fall. 

We can use NGC 6334 as a test case to estimate the fraction of the YSO population our near-infrared criteria will be able to recover when applied to other massive star forming regions. If we apply only our near-infrared selection cuts to NGC 6334, we identify 1602 (84\% of the 1908 total) Class II YSOs. Of the 375 Class I YSOs, our near-infrared selection cuts identify 189 (50\%), biased toward detecting the more evolved (less-red) sources. The statistics of the color-identified YSO candidates are summarized in Table~\ref{tab:YSOCCD}.

\subsection{SED Slope Comparison}
We have also examined the class distribution of YSOs determined by fitting the slope of the SED in the mid-infrared. The class identification for the slope of log($\lambda$F$_{\lambda}$) vs. log($\lambda$) between 2 and $\sim20$ $\mu$m from \citet{lad87} is:

\begin{equation} \alpha\geq0.3 \textnormal{ Class I}\nonumber\end{equation}
\begin{equation}0.3 >\alpha\geq-0.3 \textnormal{ Flat Spectrum}\nonumber\end{equation}
\begin{equation}-0.3 >\alpha\geq-1.6 \textnormal{ Class II}\nonumber\end{equation}
\begin{equation}-1.6 >\alpha\geq-2.7 \textnormal{ Class III}\nonumber\end{equation}

To determine the intrinsic slope of the YSO's SED (spectral index $\alpha$), we should first de-redden the observations to remove the effect of line-of-sight material. However, it is difficult to gauge how deeply embedded a source is within the molecular cloud. Using the value in the extinction map will give you the best guess for the extinction if the source was located on the far side of the molecular cloud, which could result in serious over-correction for many sources. The effect of the interstellar extinction is most prominent at short wavelength, e.g. $K_{s}$ band and shorter in this survey. We have chosen not to deredden these sources, and instead we compare how the YSO class distribution changes when we only fit the slope of the SED at longer wavelengths. We compare the distribution of $\alpha$ for the sources fit from $K_{s}$ band and longward compared to fitting only from 4.5 $\mu$m and longer, similar to the methods used in \citet{meg09} and \citet{kry12}.

\begin{figure}
\includegraphics[angle=0,width=0.9\textwidth]{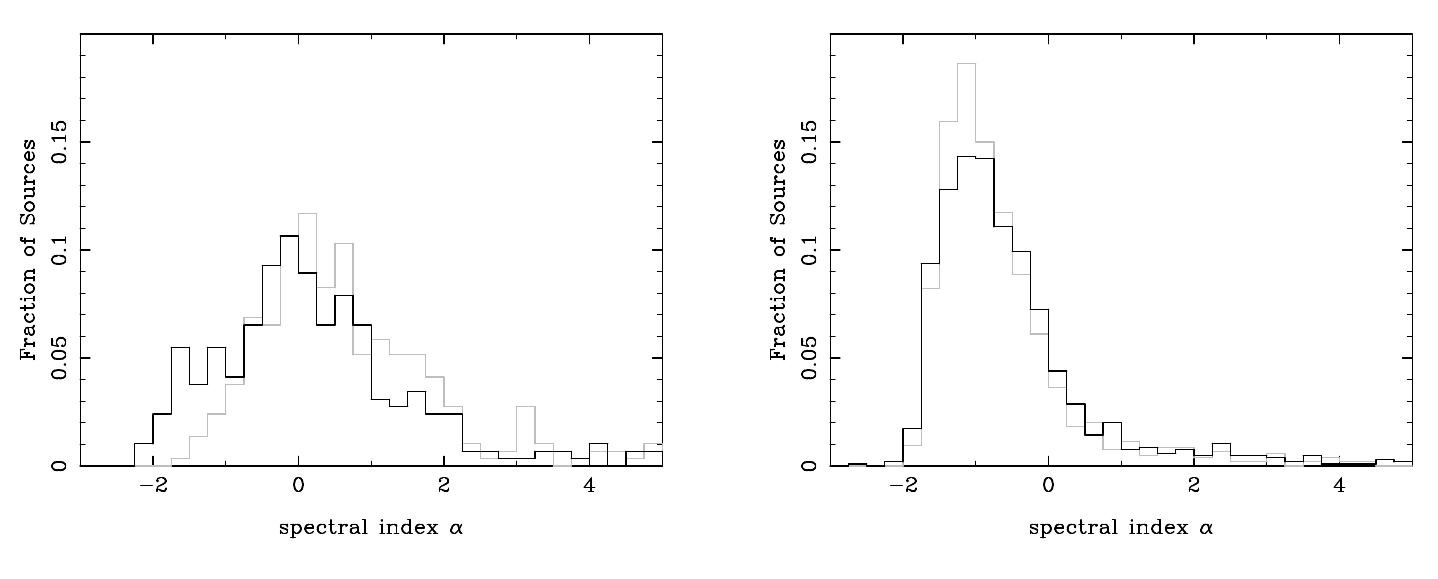}
\caption{The distribution of SED spectral index $\alpha$ for sources identified as YSO candidates using the color criteria outlined in Section~\ref{sec:YSOID}. The left panel shows Class I sources and the right panel shows Class II sources. The slope is determined by fitting the SED for at least 3 bands starting from $K_{s}$ band (gray / narrow line) or 4.5 $\mu$m (black / thick line) through 24.0 $\mu$m.}\label{fig:ColorYSOSlopeHist}
\end{figure}

Table~\ref{tab:YSOSED} shows the distribution of the \allyso{} color-identified YSO candidates by class as given by the spectral index $\alpha$. Figure~\ref{fig:ColorYSOSlopeHist} shows how the distribution of the spectral index $\alpha$ differs when using a minimum wavelength of $K_{s}$ versus 4.5 $\mu$m to fit the power-law to the SED. When we use $K_{s}$ band as the minimum wavelength to fit the spectral index $\alpha$, we see a shift in the distribution towards a greater number of Class I / Flat Spectrum sources. This likely corresponds to the difficult to separate population of Class II YSOs that are more deeply embedded in the cloud, and thus seen behind a larger amount of interstellar extinction. Source SED classification that relies solely on $\lambda \geq 4.5$ $\mu$m is therefore less sensitive to the effects of extinction and provides a superior estimate of evolutionary state. Even when fitting the SED from 4.5 $\mu$m longward, we see a number of sources identified as Class I YSOs based on the color-color diagrams that appear more likely to be Class II YSOs based on the full slope of the SED. As we will discuss in Section~\ref{sec:ExtMap}, the observed area of NGC 6334 contains a significant amount of high column density material that could contribute to cross-contamination of Class I YSOs by reddened Class II YSOs. Even using the power-law fit from 4.5 $\mu$m longward, the vast majority of color-identified YSO candidates meet the criteria to be either Class I or Class II YSOs, and all have $\alpha \geq-2.7$ indicating excess emission at infrared wavelengths indicative of circumstellar material. Table ~\ref{tab:YSOCSAMP} lists all of the YSO sources identified by the IRAC and near-infrared based color selection criteria and gives $\alpha$ fit from $K_{s}$ band longward.

\subsection{Contamination by Foreground and Background Sources in Warm Spitzer Data}\label{sec:FGBGCONT}

Misidentification of contaminant objects as YSOs remains an issue for photometry-based identification methods. Although the Gutermuth et al. criteria are very useful at identifying potential contaminant objects, they require detection at 5.8 or 8.0 $\mu$m. Now that Spitzer has exhausted its cryogen and is operating in the warm mission, new IRAC observations of star forming regions are only able to obtain photometry at 3.6 and 4.5 $\mu$m. We have used our near-infrared YSO identification criteria defined in Section~\ref{sec:NIRYSO} to quantify the amount of contamination that we would have seen in our YSO population if our observations for this region had been limited to only warm mission data. If we applied only the near-infrared criteria to our full point source catalog we would identify 1,579 sources as YSO candidates: 189 Class I YSOs (50\% complete) and 1602 Class II YSOs (84\% complete).

Background galaxies are one class of potential contaminant objects. Near-infrared galaxy counts from \citet{mar01} estimate that a survey with our sensitivity would detect approximately $1.3\times10^{4}$ galaxies per deg$^{2}$, consistent with the 10,327 suspected extragalactic sources in our catalog identified by the \citet{gut09} criteria. Of these galaxies identified by their IRAC 4 band colors, 62 have the correct colors to match our  $H$, $K_{s}$, 3.6 and 4.5 $\mu$m YSO color criteria. Only 1 of these 62 sources is red enough to match the Class I criteria; the other 61 sources are only moderately red and would have been identified as Class II YSOs. We therefore would expect that approximately 3.8\% of detected Class II YSO candidates and 0.5 \% of detected Class I YSO candidates identified using only near-IR and warm Spitzer color criteria would actually be background galaxies. 

The primary culprit for Galactic contamination of our YSO population is Asymptotic Giant Branch (AGB) stars. To asses the level of contamination by evolved stars in our warm \emph{Spiztzer} YSO classification scheme, we referenced a sample of spectroscopically confirmed, intrinsically red post-main sequence stars in the Large Magellanic Clouds, identified by various authors and matched to IRAC and near-IR photometry from the "Surveying the Agents of a Galaxy's Evolution" (SAGE, \citealt{mei06}) catalog. We used the selection criteria from Section~\ref{sec:YSOID} to estimate the fraction of post-main sequence sources matching our YSO colors. Although none of the sample RGB and early-AGB stars matched the YSO criteria, we found matches among the more evolved stars.

Our post-main sequence LMC template sample included 423 AGB stars (from \citealt{mat09} and \citealt{kon01}), 267 carbon-rich and 156 oxygen-rich. Of these, 35 AGB stars passed the selection cuts and were identified as YSO candidates, 10 Class I and 25 Class II. Out of the sample of 75 post-AGB stars from \citet{van11}, 1 matched the Class I YSO criteria and 25 match the Class II YSO criteria. We would identify 2.4\% of the AGB stars in a field as Class I YSOs and 5.9\% of the field AGB stars as Class II YSOs. One of the previous results from \emph{GLIMPSE} was a quantification of the Galactic distribution of intrinsically-red stellar sources \citep{rob08}. At $l=351\degr{}$, $b=0.7\degr{}$ \cite{rob08} determined that there would be approximately 80 AGB sources per square degree; indicating that our 0.8 square degree region covered from $J$ band through 4.5 $\mu$m should contain roughly 68 AGB stars. By applying the SAGE-derived contamination fraction estimates, we would have falsely identified 2 of those AGB stars as Class I YSOs and 4 of those AGB stars as Class II YSOs. We tabulated the fraction of different types of contaminant sources that have colors compatible to Class I and II YSOs in Table~\ref{tab:YSOCONT}. We examined a sample of 75 post-AGB stars \citep{hug90,pin10,pom08,woo11}. We found only 1.3\% of the post-AGB sources matched our near-infrared Class I YSO selection cuts, but fully one-third (33.3\%) matched the colors for Class II YSOs. Fortunately the post-AGB phase evolves rapidly ($<10^5$ years, \citealt{vanw03}) and they tend to be uniformly distributed on the sky, so the contamination of YSO samples by post-AGB stars within a star forming region should be negligible.

Final confirmation of the nature of our YSO candidates would require follow-up spectroscopic observations. Similar follow-up studies in other star forming regions have recently shown that the contamination fraction of color-selected YSO candidates can be quite high, especially for sources with small and moderate infrared excess: for transition disk objects in 8 star forming regions, the observed AGB contamination fraction ranged between 20 and 100\% \citep{rom12}. Their identified transition disk objects all had $[3.6]-[4.5]<0.25$. In our sample of color-identified Class II YSOs this cut applies to only 28 sources.

\section{DISCUSSION}\label{sec:Discussion}

The results of the YSO census for Class I and Class II objects allows us to estimate global properties of the stellar population associated with NGC 6334, such as its rate of formation and the star formation efficiency. By expanding to use the entire point source catalog we can also map the extinction across the survey area to help determine NGC 6334's cloud structure and mass.

\subsection{Extinction, Column Density and Total Gas Mass}\label{sec:ExtMap}
\begin{figure}
\includegraphics[width=0.9\textwidth]{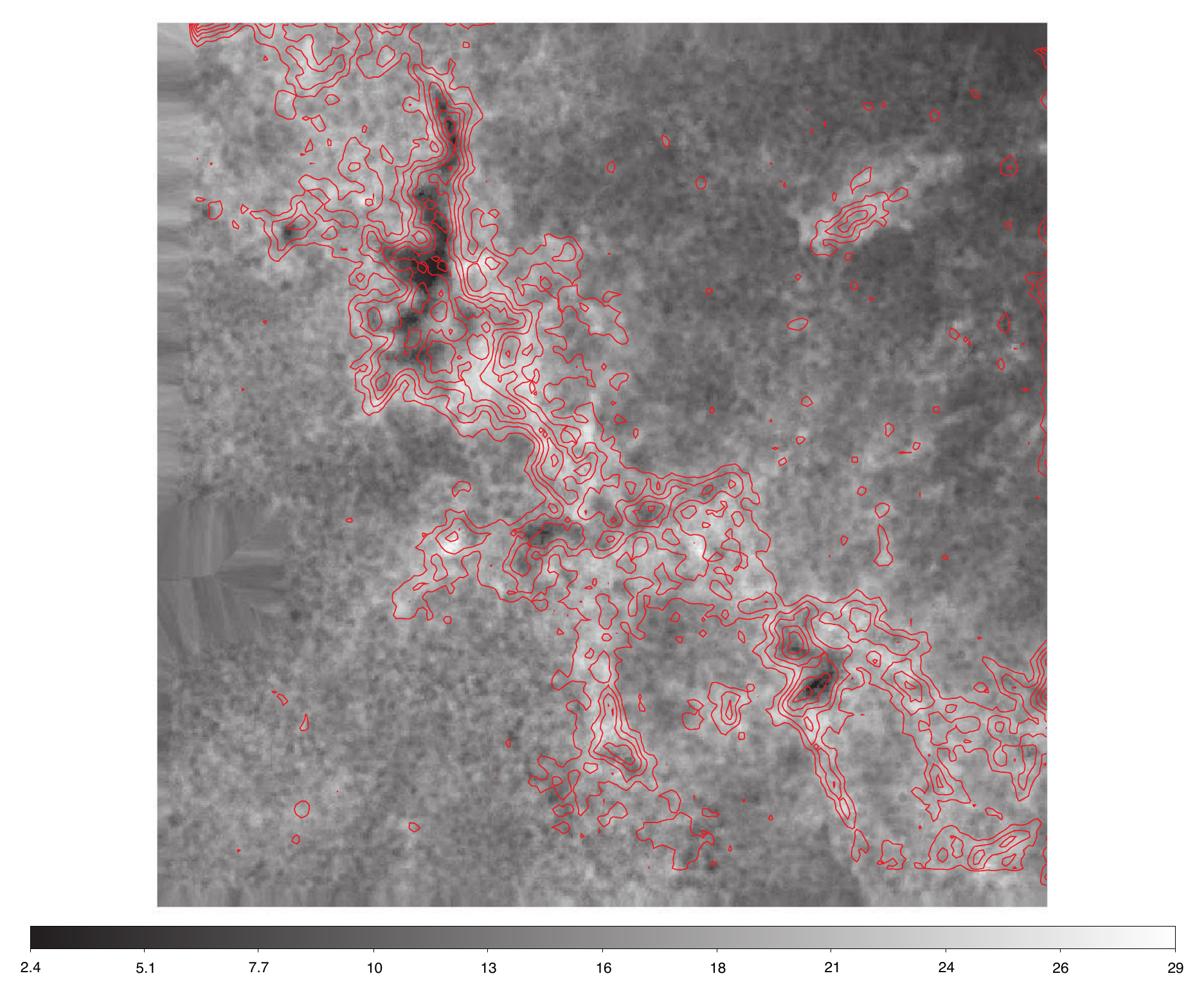}
\caption{The background shows the extinction towards NGC 6334 as determined by the NICER algorithm. The extinction map saturates at $A_{V}>30$ due to an insufficient number of background sources. The contours trace the observed density of stars, where the highest contour level corresponds to the lowest observed density of stars. The structures that correspond to NGC 6334's dark clouds, as seen in Figure~\ref{fig:IRACRGB}, exhibit anomalously low extinction values due to an almost complete lack of detection of background sources.}\label{fig:EXTMAP}
\end{figure}

The NICER algorithm \citep{lom01} was applied to the sources in the 0.8 square degree region covered at $J$, $H$, $K_{s}$, 3.6 and 4.5 $\mu$m. This method uses the $J-H$, $H-K_{s}$, $K_{s}-[3.6]$ and $[3.6]-[4.5]$ near-infrared colors to determine the line of sight extinction towards each source. The extremely dense conditions in the dark clouds in NGC 6334 result in regions lacking sufficient numbers of stars to accurately determine the true value of the extinction. These extremely dense regions have a lower bound of $A_{V}>30$ and overlap with the dark clouds of the field as seen in all 4 IRAC bands and even the MIPS 24 $\mu$m images. In these regions we used the local stellar density to derive a lower bound for the total extinction. The un-patched extinction map is shown in Figure~\ref{fig:EXTMAP}, with contours of point source density to highlight the areas where the large average distances between point sources suggest the low extinction values in the map are anomalous.

Dust column densities derived from Herschel Hi-GAL and HOBYS results show that along the dense ridges in NGC 6334, the total visual extinction exceeds several hundred magnitudes \citep{rus13}. Approximately 10 pc$^2$ along the ridge have column densities that exceed the 0.7 $g$ $cm^{-2}$ ($A_{V}\sim200$) suggested as the threshold for massive stars formation by \citet{kru08}. The extremely high column densities in the ridge are supported by extinction measurements from Chandra X-ray observations of NGC 6334 \citep{fie09}.

A recent analysis of the optically visible O and B stars in NGC 6334 determined an average line of sight extinction of $A_{V} = 4.12 \pm 0.20$ \citep{per08}. We subtracted this line of sight component from the patched extinction map and summed the extinction over the map, converting from magnitudes of extinction to column density using $N(H_{2})=1.37\times10^{21}A_{V}$ \citep{hei10,dra03}. At the assumed distance to NGC 6334 of 1.6 kpc, the 0.75 deg$^{2}$ extinction map corresponds to 590 pc$^{2}$. We assume a standard composition of 63\% Hydrogen, 36\% Helium and 1\% dust to determine the total cloud mass from the calculated hydrogen column density. We find the mass of NGC 6334 is $2.2 \times 10^{5}$ M$_{\odot}$. Due to the limitations of calculating column density from extinction when the extinction map saturates at approximately $A_{V}\sim30$, this mass should be considered a lower limit. Indeed, the HOBYS team reports $3.8\times10^{5}$ $M_{\sun}$ for NGC 6334 \citep{rus13}. However, the area surveyed by the HOBYS team also extends significantly farther than our extinction map, which may contribute to their higher mass estimate.

Our mass estimate closely matches the mass derived from the HiGAL column density map pertaining to the same area. After subtracting the line of sight component, we find up to 70\% of the mass of cloud appears to be contained in dense material above the general star formation threshold of $A_{V}\sim10$ (e.g.\citealt{mck89,lad10,hei10}). This high mass concentration is found both in the extinction map derived from NICER as well as the HiGAL column density map.

\subsection{NGC 6334 Star Formation Rate and Efficiency}\label{sec:IntRat}

In nearby star forming regions, it is reasonable to assume that the observed sample of YSOs is nearly complete and assign an average mass to the observed population in order to estimate physical characteristics such as the star formation rate and efficiency. For NGC 6334, due to the increased distance, the implementation of strict criteria to identify the YSOs and the effects of saturation and decreased efficacy of source detection in the bright nebular emission, the observed Class I and Class II populations are noticably incomplete. In addition, due to the crowding of this field we have not attempted to isolate the Class III YSO population from the Galactic foreground and background stars in our point source catalog. 

We have instead used the observed Class I and Class II sample to estimate the total size of the YSO population in NGC 6334. We have compared our work to the NGC 6334 X-ray point source catalog to estimate the number of young stars in the complex that lack significant infrared excess and to meet the YSO identification criteria \citep{fie09}.
 
\subsubsection{Estimating Stellar Population Mass}\label{sec:YSOMASS}

We have used the observed population of YSO candidates and YSO models to estimate the lowest mass objects at which our survey is complete, and infer the lower-mass population from that point by assuming a standard Kroupa IMF.

First, we took the simulated photometry for the 200,710 models computed by \citet{rob06} and scaled the observed magnitudes from 1 kpc (the default distance) to the adopted distance to NCC 6334, 1.6 kpc. We ran the model photometry through the YSO selection cuts we outlined in Section ~\ref{sec:YSOID}. This enabled us to isolate model YSO SEDs with the same observable properties as the Class I and Class II YSOs that we identified in NGC 6334. Models with relatively high envelope accretion rates were excluded as Class 0 sources that are generally not detected below 10 $\mu$m. After identifying the Class I and Class II YSO models, the leftover models that were not red enough to meet Class I or Class II selection criteria were selected to represent Class III YSOs.

We examined each YSO Class independently and grouped models by their 3.6 $\mu$m magnitude in bins 0.1 magnitudes wide. In each magnitude bin we looked at the distribution of stellar mass. We selected the peak of the distribution as the average mass for models within that magnitude range. We integrated under the stellar mass distribution around the peak until our mass range included 65\% of the models. The smallest mass within this range is adopted as the lower bound for the mass, and the highest mass within this range is adopted as the upper bound. Using this approach, we have established an average mass-magnitude relation for the YSO models based on their brightness at 3.6 $\mu$m.

\begin{figure}
\includegraphics[angle=0,width=0.9\textwidth]{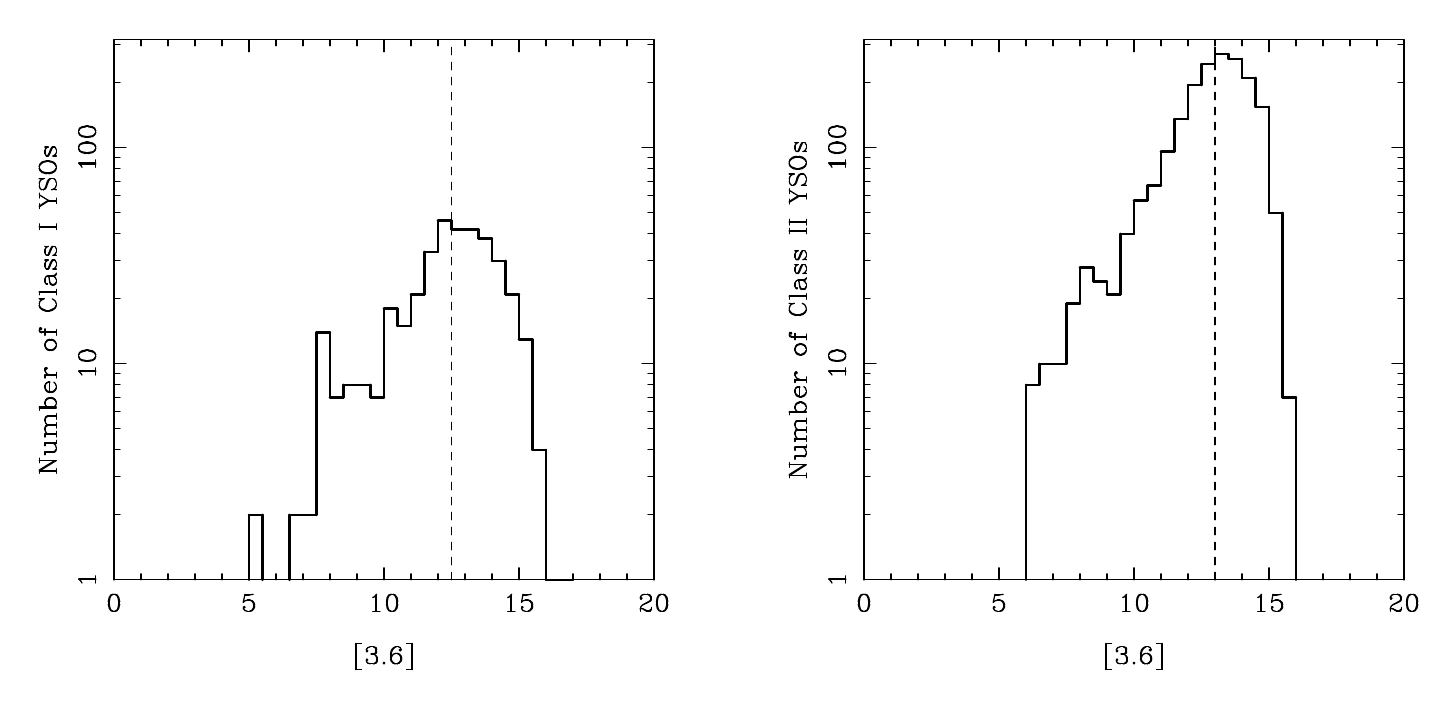}
\caption{The 3.6 $\mu$m magnitude histograms for the \allci{} Class I (left panel) and \allcii{} Class II (right panel) YSO candidates. The dashed lines mark the approximate magnitude at which the YSO census is assumed to be complete for each class, at $[3.6]=12.5$ for Class I YSOs and $[3.6]=13$ for Class II YSOs.}\label{fig:YSOLF}
\end{figure}

Next we used the observed 3.6 $\mu$m magnitude histogram for the NGC 6334 Class I and II YSOs (shown in Figure~\ref{fig:YSOLF}) to determine the approximate magnitude at which each class is complete, identified by the point at which the magnitude histogram ceases to increase. In this way, we determine the Class I YSO population is complete to $[3.6]=12.5$ with 220 YSOs brighter than this limit. The Class II YSO population is complete to $[3.6]=13$, with 987 YSOs brighter than this limit. We used the average mass-magnitude relation derived from the YSO models to translate our magnitudes into a mass range. We find the $[3.6]=12.5$ for Class I YSOs corresponds to $M_{complete}=0.75^{+0.9}_{-0.6}$ $M_{\sun}$, with a median value of 1.2 $M_{\sun}$. For Class II YSOs we find $M_{complete}=0.3^{+0.3}_{-0.15}$ $M_{\sun}$, with a median value of 0.3 $M_{\sun}$, all at the assumed distance of 1.6 kpc.

We used the number of YSOs brighter than the completeness limit (220 Class I and 987 Class II) to normalize a Kroupa IMF with $\alpha=2.3$ for $M>0.5 M_{\sun}$ and $\alpha=1.3$ for $M<0.5 M_{\sun}$. By extrapolating from $M_{complete}$ down to the hydrogen burning limit (0.08 $M_{\sun}$), we estimate that there are approximately \numimfci{} Class I YSOs in NGC 6334 and \numimfcii{} Class II YSOs. The derived mass in Class I YSOs is \mci$^{+1500}_{-600}$ $M_{\sun}$. The derived mass in Class II YSOs is \mcii$^{+1400}_{-500}$ $M_{\sun}$.

Although we did not directly isolate the Class III population in our point source catalog, we can use the fact that young stars have a high level of X-ray activity to characterize the population of the more evolved stars in NGC 6334. A Chandra X-ray survey of NGC 6334 \citep{fie09} detected approximately 1,600 sources identified as pre-main sequence stars; we find only 49 Class I and 165 Class II YSOs within 6\arcsec{} of these X-ray sources, confirming that the majority (almost 90\%) of the X-ray sources are likely to be more evolved Class III YSOs / pre-main sequence stars that are otherwise not included in our YSO census. \citet{fie09} determined that the X-ray sample was complete to approximately 1 $M_{\odot}$, and that extrapolating to low X-ray levels below their detection threshold would suggest a total of 20,000 to 30,000 pre-main sequence stars associated with the complex. 

\begin{figure}
\includegraphics[angle=0,width=0.9\textwidth]{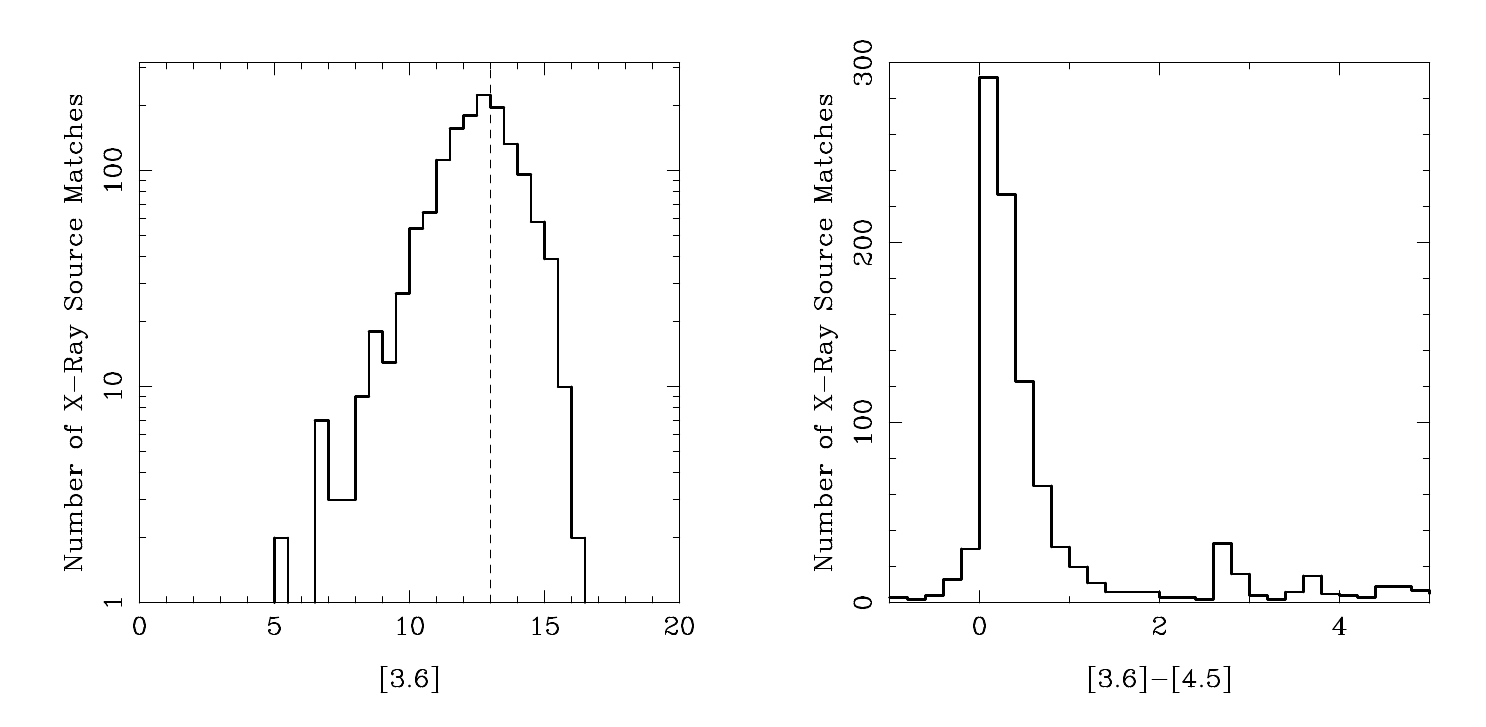}
\caption{The left panel shows the 3.6 $\mu$m magnitude historgram for the 1,408 sources adopted as the Class III population that match within 5$''$ of a \emph{Chandra} X-ray point source. The dashed line at 3.6 $\mu$m = 13 marks the approximate magnitude at which the Class III YSOs census is assumed to be complete. The right panel shows the distribution of [3.6]-[4.5] for the 1294 X-ray sources detected in both IRAC bands.}\label{fig:YSOCIIILF}
\end{figure}

We matched our IRAC sources detected at 3.6 $\mu$m to the Chandra X-ray catalog, finding 1,400 total matches and adopt this as our Class III YSO population. Many of these sources display minimal infrared excess, with a median $[3.6]-[4.5]$ of 0.09. As seen in Figure~\ref{fig:YSOCIIILF}, the Class III sources are complete to $[3.6] = 13$, with 873 sources brighter than this limit. From the Robitaille models, for Class III YSOs $[3.6]=13$ corresponds to approximately $M_{complete}=1.5\pm{0.6}$ $M_{\sun}$ at a distance of 1.6 kpc. We apply the same method of normalizing a Kroupa IMF as we applied to the Class I and Class II YSOs. We find the 873 Class III / pre-main-sequence stars detected brighter than $[3.6] = 13$ represent a total population with mass $7900^{+4700}_{-3900}$ $M_{\sun}$. Using a standard 0.5 $M_{\sun}$ average stellar mass for the 20,000 - 30,000 stars reported by \citet{fie09} we would expect 10,000 - 15,000 $M_{\sun}$ in Class III YSOs in NGC 6334. The high level of diffuse emission that makes it difficult to achieve a uniform completeness across the observed field means that even our YSO samples that are assumed to be complete may be missing a significant number of sources. The true value for the YSO stellar mass are therefore likely to be closer to the upper end of the ranges derived by examining the sources detected at 3.6 $\mu$m. By extension,the derived physical quantities we report (star formation rate and efficiency) may also under-represent the total star formation activity in NGC 6334 by a factor of up to 2.

\subsubsection{Rate and Efficiency of Star Formation in NGC 6334}

The HII regions in NGC 6334 are estimated to be on the order of 2 Myr old or younger \citep{per08}. We adopt 2 Myr as the age of young stars in NGC 6334 and use the total mass in YSOs to determine the overall star formation rate~$SFR\sim4900^{+3800}_{-2500}$~$M_{\sun}$~Myr$^{-1}$. The 0.8 deg$^{2}$ area used to derive the YSO catalog corresponds to approximately 600 pc$^2$ at a distance of 1.6 kpc. Over this wide-field view of NGC 6334, we find that the average star formation rate surface density is $8.2^{+6.3}_{-4.2}$ \unitssigsfr{}.

We use the Class I YSO population mass extrapolated from the observed Class I YSOs to estimate the current star formation rate in NGC 6334. The mean lifetime of Class I YSOs has been estimated to be $0.4\pm{}0.2$ Myr in low mass star forming regions \citep{eva09}, although this is likely to be shorter for more massive stars. The average value of our Class I total mass estimate from Section~\ref{sec:YSOMASS} is \mci{} $M_{\sun}$. We use the 0.4 Myr lifespan of Class I YSOs and find the current star formation rate in NGC 6334 is $2000^{+3700}_{-1600}$ $M_{\sun}$ Myr$^{-1}$. In NGC 6334 the Class I YSOs are almost exclusively found in regions with $A_{V}>8$, which accounts for almost 25\% of the area we have observed (150 pc$^{2}$). Within this dense gas the star formation rate surface density is $\Sigma_{SFR,CI}=13.0^{+24.7}_{-10.4}$ $M_{\sun}$ Myr$^{-1}$ pc$^{-2}$.

We determine the star formation efficiency in NGC 6334 by adding up the mass of the Class I, II, and III YSOs and comparing to the cloud mass estimate determined from the extinction map. The poorly-characterized Class 0 population may be safely ignored in the star formation efficiency computation; due to the relatively short time scales involved in Class 0 evolution, these sources would contribute only a negligible amount to the total stellar mass content of NGC 6334. Overall, this yields an integrated stellar mass content of NGC 6334 of $\mallysolow{} - \mallysohigh$~$M_{\sun}$. Adopting the estimated $2.2\times10^{5}$ $M_{\sun}$ gas mass for the cloud from Section~\ref{sec:ExtMap}, we estimate the lower limit for the efficiency of star formation activity in NGC 6334 is $SFE\sim0.04^{+0.03}_{-0.02}$. If we adopt the \citet{fie09} estimate for the size of the stellar X-ray population in NGC 6334 the efficiency may exceed the upper limit of the range we found, $SFE\ge0.10$. Among the dense cores and clumps of the cloud we find local star formation efficiency upwards of 0.3. The local variations in the efficiency of star formation and sites of cluster formation within NGC 6334 will be discussed in a followup paper.

The star formation rate and efficiency we present in this paper are derived using 5$\sigma$ as the maximum photometric error in each band for inclusion in our point source catalog. Sources with higher photometric errors intrisically have less reliable photometry and have a higher probability to be contaminant objects. We examined the number of YSO candidates we would select from our point source catalog if we changed the maximum allowed photometric error. By decreasing the maximum allowed photometric error to 0.108 (10$\sigma$) we would decrease the number of YSO candidates by approximately 25$\%$, from the original 2283 YSO candidates to 1652 YSO candidates. The spatial distribution of the 10$\sigma$ selected YSOs is virtually indistinguishable from the 5$\sigma$ sources. Additionally, despite the decrease in the number of YSO candidates we see only a small effect on the derived rate and efficiency of star formation. Our method of determining the mass in YSOs relies on the assumption that our observations are only complete down to a limited magnitude / mass range, and most of the sources observed with photometric errors between 5$\sigma$ and 10$\sigma$ fall below the completeness limit. If we exclude all sources between 5$\sigma$ and 10$\sigma$ and calculate the star formation rate surface density we would obtain $\Sigma_{SFR,10\sigma}=7.8^{+5.1}_{-4.3}$ $M_{\sun}$ Myr$^{-1}$ pc$^{-2}$, a difference of less than 10\% from the value calculated on the full set of YSO candidates detect at 5$\sigma$ or greater.

\subsubsection{Comparing NGC 6334 to Orion and Low Mass Star Forming Regions}
To verify the star formation rates calculated from this method we have applied our method of determining the star formation rate to the YSO population detected in Orion. The \emph{Spitzer} Orion survey observed a 572 pc$^{2}$ region and identified a total of 3,479 YSOs \citep{meg12}. In Orion the YSO candidates were identified using similar color-color diagrams and then the YSO candidates were primarily classified based on the slope of their SEDs. \citet{meg12} report 488 likely protostars, and the rest are considered stars with disks. \citet{gut11} reports the mass within the surveyed area of Orion is $3.3\times10^{4}$. The total mass of Orion A and Orion B is $\sim2\times10^{5} M_{\sun}$ \citep{bal08}, similar to the mass we have estimated for NGC 6334. This similarity makes Orion a good benchmark for comparison for NGC 6334.

First, we assume that the sources classified as protostars in Orion are roughly the same class as our Class I YSO population in NGC 6334, and that the stars with disks in Orion are equivalent class to the Class II YSOs in NGC 634. We estimate the Orion YSO population is complete to approximately [3.6]=14 ($M_{complete}=0.2\pm0.16$) for protostars and [3.6]=11 ($M_{complete}=0.3\pm0.4$) for stars with disks. Following the method described above, this leads to a total mass in protostars of $320^{+220}_{-150}$ $M_{\sun}$ and $1540^{+770}_{-1040}$ $M_{\sun}$ for stars with disks. Adopting the estimated 2 Myr age suggested for the Orion YSO population gives a total star formation rate $930^{+500}_{-600}$ $M_{\sun}$ Myr$^{-1}$, or a star formation rate surface density of $1.6^{+0.9}_{-1.0}$ \unitssigsfr{}. If we follow the standard assumptions for low mass star forming regions and apply them to Orion, we adopt a 0.5 $M_{\sun}$ average mass and a 2 million year star formation time scale and obtain $SFR= 870$ $M_{\sun}$ Myr$^{-1}$, within 10\% of the rate obtained using our method.

Despite the similar masses of the two clouds, the star formation rate in Orion is almost a factor of 4 smaller than the star formation rate we find in NGC 6334. One possible explanation for this is the difference in dense gas in the two clouds. In NGC 6334, we found that up to 70\% of the mass of the cloud is contained in dense gas above $A_{V}>8$. In Orion, this value is reported to be closer to 10\% \citep{lad09}. The much larger reservoir of star-forming gas suggests that NGC 6334's increased star formation rate will continue. It is important to remember that the completeness of NGC 6334's observed YSO population may be even lower than predicted here in this study.

\begin{figure}
\includegraphics[angle=0,width=0.9\textwidth]{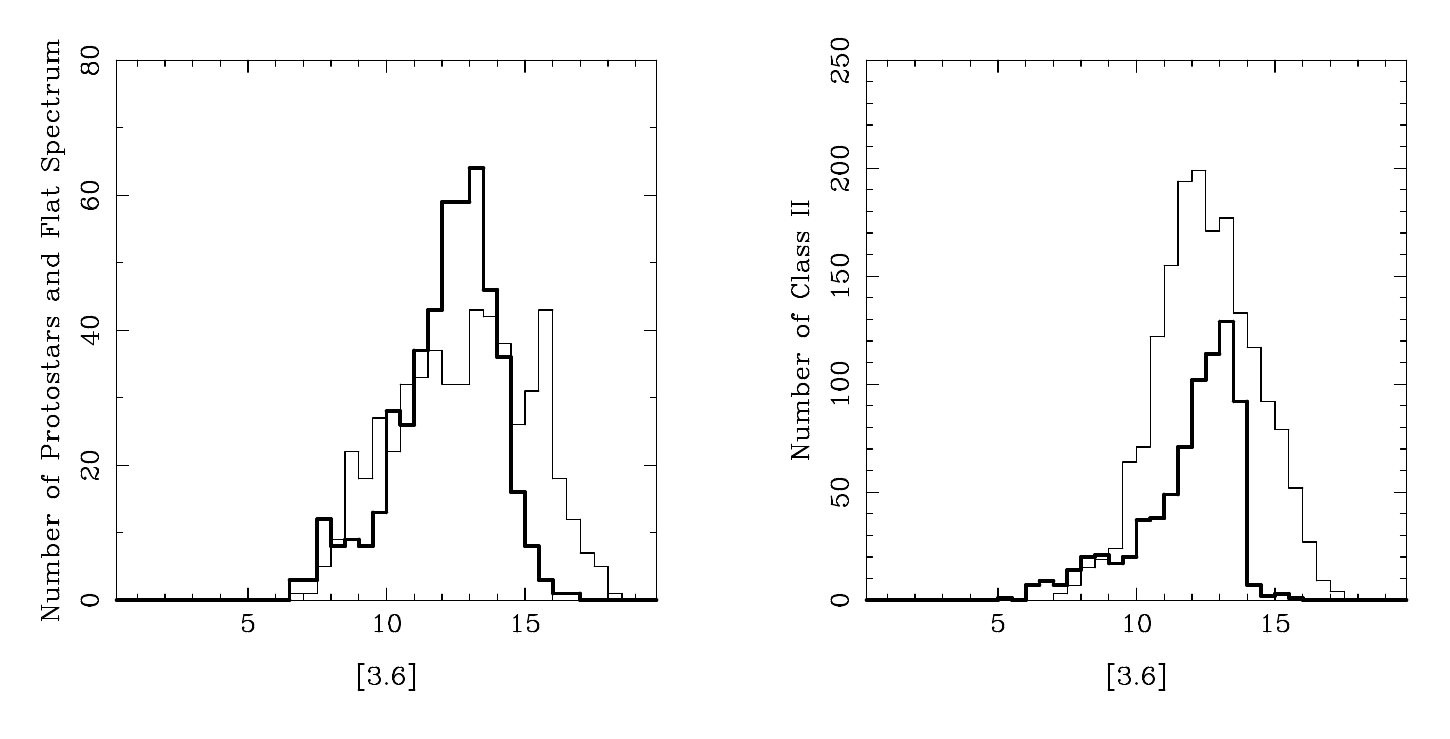}
\caption{A comparison of the 3.6 $\mu$m magnitude histograms for the YSO candidates in NGC 6334 (bold line) and Orion (thin line) with protostars / Class I YSOs in the left panel and stars with disks / Class II YSOs in the right panel. The Orion magnitudes have been shifted from an assumed distance of 400 pc to 1600 pc. The Orion survey detects sources several magnitudes fainter. Overall the bright end of the source distribution looks similar in both regions, but the distributions deviate prior to the adopted completeness limit of $[3.6]=12.5$ for the rising/flat spectrum protostars or for $[3.6]=13$ for the Class II YSOs.}\label{fig:ORIONNGC6334}
\end{figure}

In Figure~\ref{fig:ORIONNGC6334} we compare the 3.6 $\mu$m magnitude histograms of sources in NGC 6334 and Orion with the same spectral index $\alpha$. We find that the distribution of protostars and flat spectrum sources in NGC 6334 is similar on the bright end. However, the \emph{Spitzer} Orion survey extends to fainter sources as the difference in distance between the two regions would suggest. For stars with disks / Class II YSOs ($-0.3>\alpha>-1.6$) we find substantial differences between the Orion and NGC 6334 YSO 3.6 $\mu$m magnitude histograms. This difference could be attributed to intrinsic differences between the stellar populations forming in the two regions - i.e. a different shape of IMF in one region vs. another, or a different history of star formation between NGC 6334 and Orion. If we assume that the overall populations are intrinsically similar, we find that NGC 6334 is missing a substantial number of Class II sources as bright as $[3.6]=10$. This could partially be due to our more restrictive selection criteria that aim to minimize the number of contaminant objects in our YSO sample. Another issue is the wide-spread extremely bright extended PAH emission in NGC 6334. Earlier Spitzer studies (e.g. \citealt{cha08,koe08}) found that the IRAC sensitivity to YSOs could decrease by up to 3 magnitudes at 8.0 $\mu$m in fields with bright PAH emission. For this additional reason, we interpret the star formation rate and efficiency we have calculated for NGC 6334 as a conservative estimate.

Recent results for a sample of 20 local low mass star forming regions found that their star formation rate surface densities are between $0.1-3.4$ \unitssigsfr{} with an overall average of $1.2\pm0.2$ \unitssigsfr{} \citep{hei10} which is at least 5 times lower than NGC 6334. Heiderman et al. also provided an estimate of the rate of ongoing star formation in these same regions through their study of how the star formation rate density varied with increasing gas density in the clouds. Within the $A_{V}\sim8$ contour the star formation rate density derived from Class I YSOs in the low mass cloud sample is $0.03-1.3$ \unitssigsfr{} with an average of 0.5 \unitssigsfr{}, compared to the $13.0^{+24.7}_{-10.4}$ \unitssigsfr{} in NGC 6334.

The term ``mini-starburst'' has been used to describe very active massive star forming regions where conditions are similar to those observed in starburst galaxies \citep{mot03}. The suggested criteria for a ``mini-starburst'' are:
$\Sigma_{SFR}\sim$ 10 - 100 $M_{\sun}$ Myr$^{-1}$ $pc^{-2}$ and $SFE\sim0.15$ \citep{mot12}. In the IRDC G035.39-00.33 these conditions are met over an approximately 8 pc${^2}$ region, comparable in size to the dense ridge in NGC 6334. These conditions are found on a wider scale in the massive Galactic star forming region W43 located at the intersection of the Galactic bar and the Scutum-Centaurus arm. The (conservative) estimate $8.2^{+6.3}_{-4.2}$ for the global star formation rate we present in NGC 6334 is approaching the limit defining mini-starburst conditions. These conditions are fully met when we confine ourselves to the Class I YSOs located within $A_{V}>8$, where we also expect the efficiency to increase as has been found in other star forming regions \citep{gut11}. This suggests NGC 6334 may also belong to the ``mini-starburst'' class of star forming regions. 

NGC 6334 does not appear to have any special Galactic location to explain the abundance of dense gas and the subsequent extreme star formation activity taking place within this region. Low-resolution observations of transitions of CO, $^{13}$CO, CS, and NH$_{3}$ found a $\sim10$ km s$^{-1}$ velocity gradient along the main ridge indicative of a dynamic process such as large scale infall \citep{kra99}. A large scale infall or dynamic merger of multiple filamentary structures could help explain the origin of the high fraction of dense molecular gas in NGC 6334, and in turn the high star formation activity. Recent work attributes the velocity gradient along this ridge to global graviational collapse of the NGC 6334 molecular cloud \citep{zer13}.

\begin{figure}
\includegraphics[angle=0,width=0.9\textwidth]{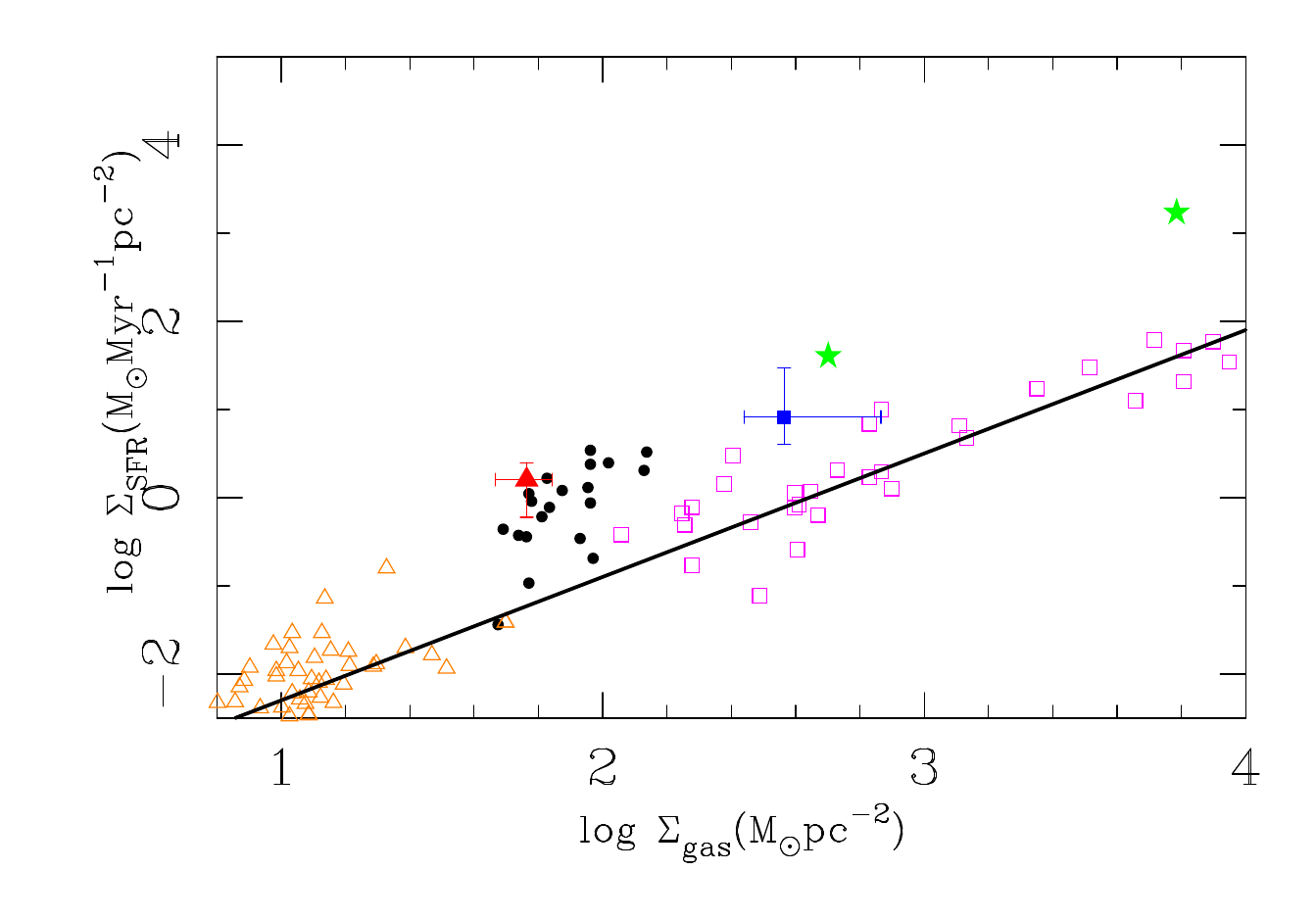}
\caption{The Kennicutt-Schmidt relation comparing the observed star formation rate surface density and gas mass surface density. NGC 6334 is the blue box and the error bars correspond to the uncertainty in the mass estimates of the YSO population and the molecular cloud. Orion is shown as a red filled triangle with corresponding errors bars at the star formation rate surface density derived with our method. Low mass star forming regions from Spitzer's c2d and GBS programs are plotted as filled black circles. W43 and G035.39, two Galactic massive star forming regions identified as prototypical $''$mini-starbursts$''$ \citep{mot12} are plotted as green stars. The open triangles show a sample of normal spiral galaxies and the open boxes show starburst galaxies. The line marks the Kennicutt-Schmidt relation with $\alpha=1.4$. The vertical offset between the extragalactic and the Galactic star forming regions likely results from the beam of extragalactic observations being diluted by non-star forming gas. A color version of this figure is available online.}\label{fig:ExGalContext}
\end{figure}

Figure~\ref{fig:ExGalContext} compares NGC 6334 to other Galactic star forming regions observed by Spitzer during the Gould Belt Survey (GBS, \citealt{har08gbs1,kir09gbs2,pet11gbs3,spe11gbs4,hat12gbs5}), the Molecular Cores to Planet-forming Disks project (c2d,\citet{eva09}), and the prototypical  ``mini-starburst'' regions W43 and G035 from \citep{mot12}. The range shown for NGC 6334 reflects the uncertainty in both the mass of the total cloud and the total age and mass of the YSO population. NGC 6334 falls between the low mass Galactic star forming regions at the lower left and the Galactic ``mini-starburst'' regions at the upper right. The location of NGC 6334 and the Galactic ``mini-starburst'' regions makes it unlikely that the vertical offset between the Galactic star forming regions and other galaxies in the Kennicutt-Schmidt diagram is due to fundamental differences between regions that are forming high and low mass stars. Instead, the high star formation rate found in NGC 6334 supports that the primary predictor of a high star formation rate is a large amount of dense gas, as suggested in e.g. \citet{gut11} and \citet{lad10}, with different dependencies on the density of the star forming gas. In a typical galaxy, the total gas mass is diluted and only a small fraction is found in molecular clouds dense enough to participate in star formation activity.

\section{SUMMARY}

We have obtained deep observations of NGC 6334 in the near- and mid-infrared, cataloging $>700,000$ Galactic point sources in a $0.8$ deg$^{2}$ region. Based on their near- and mid-infrared colors, \allyso{} of these have been identified as Class I or Class II YSOs belonging to the NGC 6334 star forming complex. Overall the observed YSO population is found to be complete to just below 1 $M_{\sun}$. We have modified existing YSO identification schemes in order to identify the population of YSOs in other star forming regions where long wavelength IRAC data is not available, as well as estimate the contamination, completeness, and reliability of those measurements. The full point source catalog was used to map the extinction across NGC 6334 to estimate the region's mass and from that the efficiency of star formation. The 0.8 deg$^{2}$ area used to derive the YSO catalog corresponds to approximately 600 pc$^2$ at a distance of 1.6 kpc. Over this wide-field view of NGC 6334, we find that the average star formation rate surface density is $8.2^{+6.3}_{-4.2}$ \unitssigsfr{}.

By fitting the observed YSOs with a Kroupa IMF we can extrapolate down to the hydrogen-burning limit to account for the rest of the stellar mass that still remains below the observation detection threshold. Combining this with the estimated pre-main sequence population from \cite{fie09} places the current global efficiency of star formation in NGC 6334 at $SFE\sim0.04^{+0.03}_{-0.02}$ based on a cloud mass of $2.2\times10^{5}$~$M_{\sun}$. The conservative estimate for the star formation rate is $8.2^{+6.3}_{-4.2}$ $M_{\sun}$ Myr$^{-1}$ pc$^{-2}$ or~$SFR\sim4900^{+3800}_{-2500}$~$M_{\sun}$~Myr$^{-1}$. Our method assumes the detection of a complete sample of YSOs down to the sensitiviy limits of our survey. The true values for the star formation rate may in fact lie closer to, or even above, the upper limits we report based on the uncertainties in the method of converting an incomplete YSO count into a mass determination.

From the estimated mass in Class I YSOs, we find the current rate of star formation in NGC 6334 is 13.0$^{+24.7}_{-10.4}$ $M_{\odot}$ Myr$^{-1}$ pc$^{-2}$ in the dense star-forming gas within the $A_{V}>8$ contour. The high efficiency and rate of star formation in NGC 6334 indicates that this region may be undergoing a burst of star forming activity, possibly characterized as a $''$mini-starburst$''$ event.

The methods of data reduction and YSO identification we have derived here will be applied to a sample of 5 other massive Galactic star forming complexes observed in a similar fashion with Spitzer IRAC and NEWFIRM. This combined sample will add to the statistical set of Galactic regions with a resolved YSO population characterizing both the high and low ranges of stellar mass. These regions can in turn be be directly compared to the Kennicutt-Schmidt relation to assist in bridging the gap between extragalactic star formation and nearby star forming regions.

\acknowledgments
We acknowledge the assistance of Robert Gutermuth, Vallia Antoniou, Andres Guzman and Rafael Martinez during the preparation of this manuscript. This work is based on observations made with the Spitzer Space Telescope, which is operated by the Jet Propulsion Laboratory, California Institute of Technology under NASA contract 1407. S.W. acknowledges partial support from NASA Grants NNX12AI55G and NNX10AD68G, and JPL-RSA 1369565. 

Facilities: Spitzer (IRAC, MIPS)

\clearpage
\begin{deluxetable}{cccccccccccc}
 \tabletypesize{\tiny}
\rotate
\tablecaption{NGC 6334 Point Source Catalog\label{tab:PSCSAMP}}
\tablewidth{0pt}
\tablehead{\colhead{Catalog Entry} & \colhead{RA (2000)} & \colhead{DEC (2000)} & \colhead{$J$} & \colhead{$H$} & \colhead{$K_{s}$} & \colhead{[3.6]} & \colhead{[4.5]} & \colhead{[5.8]} & \colhead{[8.0]} & \colhead{[24.0]} & \colhead{Classification\tablenotemark{a}}}
\startdata
7922 & 17:17:57.252 & -36:20:04.06 & 13.628$\pm{}$0.013 & 12.858$\pm{}$0.135 & 10.761$\pm{}$0.110 & 6.903$\pm{}$0.021 & 6.644$\pm{}$0.026 & 6.175$\pm{}$0.012 & 5.644$\pm{}$0.016 & 3.986$\pm{}$0.018 & ICII \\
24056 & 17:19:13.033 & -36:20:26.63 & 19.421$\pm{}$0.071 & 16.508$\pm{}$0.048 & 14.389$\pm{}$0.040 & 10.444$\pm{}$0.020 & 8.723$\pm{}$0.014 & 7.076$\pm{}$0.015 & 5.948$\pm{}$0.022 & 2.302$\pm{}$0.011 & ICI \\
26818 & 17:19:09.717 & -36:19:42.10 & 19.000$\pm{}$0.060 & 14.367$\pm{}$0.020 & 11.194$\pm{}$0.016 & 7.525$\pm{}$0.019 & 6.449$\pm{}$0.017 & 5.383$\pm{}$0.014 & 4.584$\pm{}$0.014 & 2.538$\pm{}$0.011 & ICI \\
41234 & 17:19:11.658 & -36:17:47.73 & 15.711$\pm{}$0.010 & 12.014$\pm{}$0.021 & 11.109$\pm{}$0.080 & 7.674$\pm{}$0.025 & 6.921$\pm{}$0.017 & 6.182$\pm{}$0.016 & 5.854$\pm{}$0.016 & 4.394$\pm{}$0.029 & ICI \\
52364 & 17:18:03.991 & -36:12:50.20 & 13.944$\pm{}$0.010 & 12.380$\pm{}$0.028 & 11.529$\pm{}$0.020 & 10.280$\pm{}$0.024 & 9.848$\pm{}$0.022 & 9.211$\pm{}$0.024 & 8.384$\pm{}$0.032 & 5.541$\pm{}$0.057 & ICII \\
55472 & 17:20:13.901 & -36:19:15.21 & 15.678$\pm{}$0.010 & 12.100$\pm{}$0.043 & 11.262$\pm{}$0.180 & 8.510$\pm{}$0.026 & 8.609$\pm{}$0.024 & 8.059$\pm{}$0.014 & 8.049$\pm{}$0.017 & 6.874$\pm{}$0.137 & UC \\
62902 & 17:18:00.601 & -36:11:24.54 & 15.878$\pm{}$0.009 & 13.726$\pm{}$0.006 & 12.543$\pm{}$0.010 & 11.098$\pm{}$0.024 & 10.536$\pm{}$0.025 & 9.878$\pm{}$0.026 & 8.940$\pm{}$0.020 & 5.424$\pm{}$0.044 & ICII \\
67569 & 17:17:56.043 & -36:10:49.07 & 14.958$\pm{}$0.009 & 12.994$\pm{}$0.017 & 11.610$\pm{}$0.014 & 9.614$\pm{}$0.028 & 8.852$\pm{}$0.018 & 7.987$\pm{}$0.015 & 7.071$\pm{}$0.024 & 3.759$\pm{}$0.016 & ICI \\
68855 & 17:20:28.990 & -36:18:46.29 & 16.302$\pm{}$0.010 & 12.012$\pm{}$0.099 & 10.408$\pm{}$0.029 & 8.857$\pm{}$0.019 & 8.831$\pm{}$0.026 & 8.421$\pm{}$0.017 & 8.385$\pm{}$0.023 & 6.900$\pm{}$0.137 & UC \\
86342 & 17:19:36.644 & -36:14:22.32 & 15.793$\pm{}$0.030 & 13.446$\pm{}$0.019 & 11.572$\pm{}$0.015 & 8.763$\pm{}$0.025 & 7.904$\pm{}$0.015 & 6.862$\pm{}$0.020 & 5.360$\pm{}$0.020 & 1.908$\pm{}$0.009 & ICI \\
91471 & 17:19:07.249 & -36:12:25.72 & 16.159$\pm{}$0.042 & 13.570$\pm{}$0.034 & 12.287$\pm{}$0.030 & 10.719$\pm{}$0.070 & 10.112$\pm{}$0.042 & 9.374$\pm{}$0.022 & 8.578$\pm{}$0.027 & 4.963$\pm{}$0.032 & ICII \\
95907 & 17:21:34.429 & -36:19:43.48 & 16.086$\pm{}$0.010 & 12.536$\pm{}$0.039 & 10.299$\pm{}$0.021 & 7.574$\pm{}$0.022 & 7.356$\pm{}$0.016 & 6.784$\pm{}$0.012 & 6.477$\pm{}$0.022 & 4.602$\pm{}$0.024 & UC \\
97497 & 17:20:06.286 & -36:15:06.71 & 15.086$\pm{}$0.020 & 13.246$\pm{}$0.210 & 11.333$\pm{}$0.164 & 7.970$\pm{}$0.021 & 8.008$\pm{}$0.017 & 7.459$\pm{}$0.015 & 7.438$\pm{}$0.016 & 5.836$\pm{}$0.068 & UC \\
101731 & 17:22:00.096 & -36:20:30.11 & 15.330$\pm{}$0.026 & 12.057$\pm{}$0.055 & 10.475$\pm{}$0.040 & 9.029$\pm{}$0.030 & 9.077$\pm{}$0.036 & 8.534$\pm{}$0.022 & 8.551$\pm{}$0.019 & 6.724$\pm{}$0.116 & UC \\
103075 & 17:20:08.806 & -36:14:41.09 & 18.888$\pm{}$0.040 & 13.611$\pm{}$0.030 & 10.647$\pm{}$0.030 & 7.852$\pm{}$0.026 & 7.596$\pm{}$0.020 & 6.922$\pm{}$0.017 & 6.658$\pm{}$0.020 & 4.541$\pm{}$0.027 & ICII \\
114602 & 17:21:01.598 & -36:16:30.22 & 16.651$\pm{}$0.013 & 14.709$\pm{}$0.042 & 13.424$\pm{}$0.015 & 11.546$\pm{}$0.024 & 11.153$\pm{}$0.023 & 10.561$\pm{}$0.015 & 9.944$\pm{}$0.024 & 7.098$\pm{}$0.167 & ICII \\
115364 & 17:22:02.843 & -36:19:32.04 & 13.099$\pm{}$0.114 & 12.761$\pm{}$0.121 & 11.123$\pm{}$0.080 & 7.606$\pm{}$0.024 & 7.850$\pm{}$0.017 & 7.497$\pm{}$0.022 & 7.522$\pm{}$0.022 & 7.328$\pm{}$0.204 & UC \\
117424 & 17:21:16.145 & -36:17:02.96 & 17.746$\pm{}$0.038 & 15.153$\pm{}$0.041 & 13.583$\pm{}$0.010 & 10.251$\pm{}$0.032 & 9.821$\pm{}$0.022 & 9.020$\pm{}$0.016 & 8.214$\pm{}$0.016 & 4.111$\pm{}$0.017 & ICII \\
117803 & 17:21:45.612 & -36:18:36.68 & 15.165$\pm{}$0.010 & 12.144$\pm{}$0.047 & 10.514$\pm{}$0.020 & 9.089$\pm{}$0.027 & 9.196$\pm{}$0.023 & 8.698$\pm{}$0.014 & 8.605$\pm{}$0.017 & 7.160$\pm{}$0.175 & UC \\
331879 & 17:18:03.871 & -36:20:58.77 & 16.661$\pm{}$0.014 & 13.573$\pm{}$0.020 & 12.170$\pm{}$0.026 & 10.964$\pm{}$0.060 & 11.143$\pm{}$0.058 & 10.604$\pm{}$0.030 & 10.306$\pm{}$0.086 & \nodata \nodata& PAHC \\
334274 & 17:20:58.800 & -36:15:10.93 & 14.737$\pm{}$0.010 & 12.844$\pm{}$0.045 & 11.835$\pm{}$0.015 & 11.140$\pm{}$0.029 & 6.658$\pm{}$0.001 & 11.112$\pm{}$0.018 & 10.938$\pm{}$0.034 & \nodata \nodata& SK \\
356597 & 17:21:34.755 & -35:32:51.49 & \nodata \nodata& 15.858$\pm{}$0.087 & 14.831$\pm{}$0.030 & 12.137$\pm{}$0.111 & 11.241$\pm{}$0.160 & 8.904$\pm{}$0.058 & 6.874$\pm{}$0.079 & \nodata \nodata& NCI \\
368283 & 17:21:47.940 & -35:34:23.59 & 16.772$\pm{}$0.010 & 14.866$\pm{}$0.026 & 13.894$\pm{}$0.011 & 12.468$\pm{}$0.023 & 11.990$\pm{}$0.033 & 10.779$\pm{}$0.050 & 9.648$\pm{}$0.104 & \nodata \nodata& NCII \\
370373 & 17:19:22.788 & -35:43:29.94 & 14.500$\pm{}$0.010 & 13.339$\pm{}$0.019 & 12.872$\pm{}$0.020 & 12.520$\pm{}$0.026 & 12.591$\pm{}$0.026 & 12.041$\pm{}$0.048 & 10.772$\pm{}$0.089 & 6.375$\pm{}$0.099 & PAHG \\
489995 & 17:18:47.628 & -36:17:48.77 & 20.242$\pm{}$0.099 & 16.877$\pm{}$0.023 & 15.254$\pm{}$0.020 & 14.108$\pm{}$0.038 & 13.990$\pm{}$0.039 & 13.147$\pm{}$0.088 & 12.171$\pm{}$0.135 & \nodata \nodata& AGN \\

\enddata
\tablenotetext{a}{ICI = IRAC-selected Class I YSO, ICII = IRAC-selected Class II YSO, NCI = Near-Infrared-selected Class I YSO, NCII = Near-Infrared-selected Class II YSO, PAHG= PAH Galaxy, AGN = Active Galactic Nuclei, PAHC = PAH Contaminated Aperture, SK = Shock Knot, UC = Unclassified Source}
\end{deluxetable}

\clearpage
\begin{deluxetable}{cccc}
 \tabletypesize{\scriptsize}
\tablecaption{Excerpt NGC 6334 Point Source Catalog Summary\label{tab:CatSumFull}}
\tablewidth{0pt}
\tablehead{\colhead{Band} & \colhead{Number of Sources} & \colhead{Magnitude Complete} & \colhead{Limiting Magnitude}}
\startdata
        $J$    & 495,419            & 18.75              & 21.0               \\ 
        $H$    & 568,258            & 16.5               & 19.5               \\ 
        $K_{s}$    & 412,838            & 15.0               & 17.5               \\ 
        3.6 $\mu$m  & 393,158            & 14.25              & 17.5               \\ 
        4.5  $\mu$m & 389,312            & 14.0               & 16.5               \\ 
        5.8 $\mu$m  & 173,155            & 13.0               & 15.5               \\ 
        8.0 $\mu$m  & 160,947            & 12.5               & 14.5               \\ 
        24.0 $\mu$m & 991               & 6.0                & 9.2                
\enddata
\end{deluxetable}

 \clearpage
 \begin{deluxetable}{ccc}
 \tabletypesize{\scriptsize}
 \tablecaption{YSO Candidates Identified Using Color Criteria\label{tab:YSOCCD}}
 \tablewidth{0pt}
 \tablehead{\colhead{Required Bands} & \colhead{Class I}  & \colhead{Class II}} 
 \startdata
         [3.6],[4.5],[5.8],[8.0] & \ici{} & \icii{}     \\ 
         $H$, $K_{s}$, [3.6], [4.5] & \nci{} & \ncii{}       \\ 
 \tableline
 	Total & \allci{} & \allcii{}     \\
\enddata
\end{deluxetable}

\clearpage
\begin{deluxetable}{cccccc}
\tabletypesize{\scriptsize}
\rotate
\tablecaption{SED Based YSO Classification for YSO Candidates Identified by Color Selection\label{tab:YSOSED}} 
\tablewidth{0pt}
\tablehead{\colhead{Shortest Included Wavelength} & \colhead{Total Sources} & \colhead{Class I} & \colhead{Flat Spectrum} & \colhead{Class II} & \colhead{Class III}} 
\startdata
 	$K_{s}$ & 2283 & 697 & 525 & 934 & 127 \\
 		&      & 30.5\% & 23.0\% & 40.9\% & 5.6\% \\
&&&&& \\
 	4.5 $\mu$m  & 1338 & 260 & 223 & 761 & 94 \\
		    &      & 19.4\% & 16.7\% & 56.9\% & 7.0\% \\
\enddata
\end{deluxetable}

\clearpage
\begin{deluxetable}{ccccccccccccc}
 \tabletypesize{\tiny}
\rotate
\tablecaption{Excerpt from NGC 6334 Young Stellar Objects Catalog\label{tab:YSOCSAMP}}
\tablewidth{0pt}
\tablehead{\colhead{Catalog Entry} & \colhead{RA (2000)} & \colhead{DEC (2000)} & \colhead{$J$} & \colhead{$H$} & \colhead{$K_{s}$} & \colhead{[3.6]} & \colhead{[4.5]} & \colhead{[5.8]} & \colhead{[8.0]} & \colhead{[24.0]} & \colhead{Classification\tablenotemark{a}} & \colhead{$\alpha$\tablenotemark{b}}}
\startdata
315684 & 17:18:06.496 & -36:10:25.90 & 18.398$\pm{}$0.020 & 17.064$\pm{}$0.028 & 15.870$\pm{}$0.030 & 13.905$\pm{}$0.032 & 13.466$\pm{}$0.029 & 13.445$\pm{}$0.129 &nan$\pm{}$nan & nan$\pm{}$nan & NIRCII & -1.709 \\ 
315692 & 17:21:41.687 & -36:08:58.60 & 18.946$\pm{}$0.050 & 17.222$\pm{}$0.049 & 16.037$\pm{}$0.035 & 14.438$\pm{}$0.120 & 13.761$\pm{}$0.099 & nan$\pm{}$nan & 11.777$\pm{}$0.104 & nan$\pm{}$nan & NIRCII & 0.284 \\ 
315739 & 17:18:25.276 & -35:52:59.87 & nan$\pm{}$nan &nan$\pm{}$nan &nan$\pm{}$nan & 8.333$\pm{}$0.015 & 7.589$\pm{}$0.018 & 6.826$\pm{}$0.015 & 6.484$\pm{}$0.034 & 4.938$\pm{}$0.029 & ICI & -1.320 \\ 
315741 & 17:21:05.007 & -35:45:40.96 &nan$\pm{}$nan &nan$\pm{}$nan & 16.361$\pm{}$0.053 & 13.367$\pm{}$0.029 & 12.155$\pm{}$0.023 & 10.862$\pm{}$0.020 & 9.972$\pm{}$0.036 &nan$\pm{}$nan & ICI & 1.218 \\ 
315757 & 17:20:17.905 & -36:01:25.02 & 16.458$\pm{}$0.010 & 14.574$\pm{}$0.020 & 13.637$\pm{}$0.020 & 12.437$\pm{}$0.048 & 12.158$\pm{}$0.052 & 11.793$\pm{}$0.061 & 11$\pm{}$0.087 &nan$\pm{}$nan & ICII & -1.242 \\ 
315774 & 17:21:18.731 & -35:43:11.47 &nan$\pm{}$nan & 18.574$\pm{}$0.111 & 16.747$\pm{}$0.109 & 14.332$\pm{}$0.103 & 13.867$\pm{}$0.113 &nan$\pm{}$nan &nan$\pm{}$nan &nan$\pm{}$nan & NIRCII & 0.984 \\ 
315779 & 17:19:08.870 & -35:57:49.40 & 15.911$\pm{}$0.010 & 14.460$\pm{}$0.016 & 13.564$\pm{}$0.020 & 12.375$\pm{}$0.039 & 12.058$\pm{}$0.039 & 11.73$\pm{}$0.023 & 11.057$\pm{}$0.028 &nan$\pm{}$nan & ICII & -1.312 \\ 
315783 & 17:20:55.782 & -36:04:07.13 & 18.488$\pm{}$0.070 & 15.649$\pm{}$0.030 & 13.455$\pm{}$0.020 & 9.42$\pm{}$0.018 & 7.936$\pm{}$0.020 & 6.591$\pm{}$0.014 & 5.561$\pm{}$0.015 &nan$\pm{}$nan & ICI & 1.612 \\ 
315805 & 17:20:25.945 & -35:50:47.92 &nan$\pm{}$nan & 17.258$\pm{}$0.030 & 14.586$\pm{}$0.020 & 12.043$\pm{}$0.047 & 11.5$\pm{}$0.021 &nan$\pm{}$nan &nan$\pm{}$nan &nan$\pm{}$nan & NIRCII & 1.169 \\ 
315808 & 17:22:08.859 & -35:41:17.49 & 14.302$\pm{}$0.007 & 12.926$\pm{}$0.020 & 12.048$\pm{}$0.020 & 10.457$\pm{}$0.035 & 10.134$\pm{}$0.025 & 9.769$\pm{}$0.030 & 9.399$\pm{}$0.035 &nan$\pm{}$nan & ICII & -1.601 \\ 
315819 & 17:21:09.213 & -35:49:28.93 & 20.375$\pm{}$0.139 & 14.849$\pm{}$0.010 & 11.899$\pm{}$0.020 & 9.677$\pm{}$0.020 & 9.176$\pm{}$0.016 & 8.618$\pm{}$0.015 & 8.481$\pm{}$0.016 &nan$\pm{}$nan & ICII & -1.439 \\ 
315820 & 17:19:26.314 & -36:10:16.40 &nan$\pm{}$nan & 17.370$\pm{}$0.030 & 14.530$\pm{}$0.020 & 12.811$\pm{}$0.033 & 12.494$\pm{}$0.031 & 12.092$\pm{}$0.039 & 11.282$\pm{}$0.085 &nan$\pm{}$nan & ICII & -1.176 \\ 
315826 & 17:20:22.109 & -35:50:53.84 & 17.830$\pm{}$0.020 & 15.214$\pm{}$0.020 & 13.088$\pm{}$0.020 & 10.129$\pm{}$0.023 & 9.412$\pm{}$0.017 & 8.547$\pm{}$0.012 & 7.559$\pm{}$0.029 &nan$\pm{}$nan & ICI & 0.176 \\ 
315827 & 17:20:23.518 & -35:45:17.24 &nan$\pm{}$nan &nan$\pm{}$nan &nan$\pm{}$nan & 14.709$\pm{}$0.155 & 12.837$\pm{}$0.137 & 10.614$\pm{}$0.100 & 8.29$\pm{}$0.121 &nan$\pm{}$nan & ICI & 4.658 \\ 
315842 & 17:19:13.351 & -36:09:16.10 &nan$\pm{}$nan & 16.673$\pm{}$0.030 & 12.755$\pm{}$0.019 & 10.146$\pm{}$0.026 & 9.909$\pm{}$0.018 & 9.309$\pm{}$0.022 & 9.234$\pm{}$0.018 &nan$\pm{}$nan & ICII & -1.710 \\ 
315845 & 17:22:04.564 & -36:19:52.04 & 19.029$\pm{}$0.047 & 17.134$\pm{}$0.080 & 16.525$\pm{}$0.130 & 15.069$\pm{}$0.185 & 14.431$\pm{}$0.132 &nan$\pm{}$nan &nan$\pm{}$nan &nan$\pm{}$nan & NIRCII & -0.143 \\ 
315880 & 17:19:04.841 & -36:10:05.06 & 19.437$\pm{}$0.100 & 17.934$\pm{}$0.060 & 15.730$\pm{}$0.039 & 11.352$\pm{}$0.082 & 10.662$\pm{}$0.053 & 8.329$\pm{}$0.094 & 8.348$\pm{}$0.158 &nan$\pm{}$nan & NIRCI & 1.389 \\ 
315896 & 17:21:24.899 & -35:44:22.73 & 18.423$\pm{}$0.030 & 15.640$\pm{}$0.020 & 14.319$\pm{}$0.020 & 13.484$\pm{}$0.033 & 13.162$\pm{}$0.033 & 12.898$\pm{}$0.095 & 12.186$\pm{}$0.162 &nan$\pm{}$nan & ICII & -1.469 \\ 
315897 & 17:19:04.386 & -36:09:24.02 &nan$\pm{}$nan & 18.807$\pm{}$0.135 & 16.510$\pm{}$0.079 & 14.225$\pm{}$0.052 & 13.928$\pm{}$0.046 &nan$\pm{}$nan &nan$\pm{}$nan &nan$\pm{}$nan & NIRCII & 0.506 \\ 
\enddata
\tablenotetext{a}{ICI = IRAC-selected Class I YSO, ICII = IRAC-selected Class II YSO, NCI = Near-Infrared-selected Class I YSO, NCII = Near-Infrared-selected Class II YSO}
\tablenotetext{b}{$\alpha$ fit from $K_{s}$ band longward.}
\end{deluxetable}

 \clearpage
 \begin{deluxetable}{cccc}
 \tabletypesize{\scriptsize}
 \tablecaption{Fraction of Galactic and Extra-galactic Sources Matching Warm Spitzer YSO Classification Criteria\label{tab:YSOCONT}} 
 \tablewidth{0pt}
 \tablehead{\colhead{Source Type} & \colhead{Class I} & \colhead{ClassII}} 
 \startdata
	RGB / Early AGB & - & - \\
	AGB  & 2.4\% & 5.9\% \\
	Post-AGB & 1.3\% & 33.3\% \\
	Galaxies & $<0.1$\% & 0.6\% \\ 
 \enddata
 \end{deluxetable}

 \clearpage
 \begin{deluxetable}{ccc}
 \tabletypesize{\scriptsize}
 \tablecaption{Number of Expected Galactic and Extra-galactic Contaminants Per Square Degree\label{tab:YSOCONTPERDEG}} 
 \tablewidth{0pt}
 \tablehead{\colhead{Source Type} & \colhead{Class I Contaminants} & \colhead{Class II Contaminants}} 
 \startdata
	Post-Main Sequence Stars & 2.7 & 6.4 \\
	Galaxies & - & 212.7 \\ 
 \enddata
 \end{deluxetable}

\clearpage

\clearpage

\clearpage

\clearpage

\clearpage

\clearpage

\clearpage

\clearpage

\clearpage

\clearpage

\clearpage

\clearpage

\clearpage

\clearpage

\clearpage

\clearpage

\clearpage

\clearpage

\clearpage


\begin{thebibliography}{}
\bibitem[Bally(2008)]{bal08} Bally, J.\ 2008, Handbook of Star Forming Regions, Volume I, 459
\bibitem[Benjamin et al.(2003)]{ben03} Benjamin, R. A., Churchwell, E., Babler, B. L., et al.\ 2003, \pasp, 115, 953 
\bibitem[Brogan et al.(2009)]{bro09} Brogan, C. L., Hunter, T. R., Cyganowski, C. J., Indebetouw, R., Beuther, H., Menten, K. M., Thorwith, S., 2009, \apj{}, 707, 1
\bibitem[Carey et al.(2009)]{car09} Carey, S., et al. 2009, \pasp, 121, 76 
\bibitem[Caratti O Garatti et al.(2012)]{gar12} Caratti O Garatti, A., Garcia Lopez, R., Antoniucci, S., Nisini, B., Giannini, T., Eisloffel, J., Ray, T.P., Lorenzetti, D., Dabrit, S. 2012 \aap{} 538
\bibitem[Chavarr{\'{\i}}a et al.(2008)]{cha08} Chavarr{\'{\i}}a, L.~A., Allen, L.~E., Hora, J.~L., Brunt, C.~M., \& Fazio, G.~G.\ 2008, \apj, 682, 445  
\bibitem[Churchwell et al.(2009)]{chu09} Churchwell, E., Babler, B. L. , Meade, M. R., et al.\ 2009, \pasp, 121, 213 
\bibitem[Dickinson \& Valdes(2009)]{dic09} Dickinson, M., Valdes, F. A. 2009, Guide to NEWFIRM Data Reduction with IRAF, NOAO SDM PL017
\bibitem[Draine(2003)]{dra03} Draine, B.~T.\ 2003, \araa, 41, 241 
\bibitem[Evans et al.(2009)]{eva09} Evans, N. J., II et al. 2009 \apjs{} 181, 321
\bibitem[Fazio et al.(2004)]{faz04} Fazio G. G., et al. 2004, \apjs{} 154, 10
\bibitem[Feigelson et al.(2009)]{fie09} Feigelson, E. D., Martin, A. L, McNeill, C. J., Broos, P. S., Garmire, G. P. 2009, \aj{} 132, 227
\bibitem[Fischer et al.(1982)]{fis82} Fischer, J., Joyce, R.~R., Simon, M., \& Simon, T.\ 1982, \apj, 258, 165
\bibitem[Gezari(1982)]{gez82} Gezari, D.~Y.\ 1982, \apjl, 259, L29 
\bibitem[Gutermuth et al.(2004)]{gut04} Gutermuth, R. A., Megeath, S. T., Muzerolle, J., Allen, L. E., Pipher, J. L., Myers, P. C., \& Fazio, G. G. 2004, \apjs{} 154, 374 
\bibitem[Gutermuth et al.(2009)]{gut09} Gutermuth, R. A., Megeath, S. T., Myers, P. C., Allen, L. E., Pipher, J. L., Fazio, G.G. 2009, \apj{} 184, 18
\bibitem[Gutermuth et al.(2011)]{gut11} Gutermuth, R.~A., Pipher, J.~L., Megeath, S.~T., et al.\ 2011, \apj, 739, 84
\bibitem[Harvey et al.(2008)]{har08gbs1} Harvey, P.~M., Huard, T.~L., J{\o}rgensen, J.~K., et al.\ 2008, \apj, 680, 495 
\bibitem[Hatchell et al.(2012)]{hat12gbs5} Hatchell, J., Terebey, S., Huard, T., et al.\ 2012, \apj, 754, 104
\bibitem[Heiderman et al.(2010)]{hei10} Heiderman, A., Evans, N. J., Allen, L.E., Huard, T., Heyer, M. 2010, \apj{} 723, 1019 
\bibitem[Hughes \& Wood(1990)]{hug90} Hughes, S.~M.~G., \& Wood, P.~R.\ 1990, \aj, 99, 784
\bibitem[Kirk et al.(2009)]{kir09gbs2} Kirk, J.~M., Ward-Thompson, D., Di Francesco, J., et al.\ 2009, \apjs, 185, 198 
\bibitem[Kraemer \& Jackson(1999)]{kra99} Kraemer, K.~E., \& Jackson, J.~M.\ 1999, \apjs, 124, 439
\bibitem[Kryukova et al.(2012)]{kry12} Kryukova, E., Megeath, S. T., Gutermuth, R. A., Pipher, J., Allen, T. S., Allen, L. E., Myers, P. C., Muzerolle, J., 2012, \aj{} 144, 31
\bibitem[Kennicutt(1998)]{ken98} Kennicutt, R.~C., Jr.\ 1998, \apj, 498, 541
\bibitem[Koenig et al.(2008)]{koe08} Koenig, X.~P., Allen, L.~E., Gutermuth, R.~A., et al.\ 2008, \apj, 688, 1142
\bibitem[Kontizas et al.(2001)]{kon01} Kontizas, E., Dapergolas, A., Morgan, D.~H., \& Kontizas, M.\ 2001, \aap, 369, 932
\bibitem[Krumholz \& McKee(2008)]{kru08} Krumholz, M. R. \& McKee, C. E., 2008, \nat{} 451, 1082
\bibitem[Lada(1987)]{lad87} Lada, C.J., 1987, in IAU symp. 115, Star Forming Regions, ed. M. Peimbert \& J. Jugaku (Dordrecht:Reidel), 1
\bibitem[Lada \& Lada(2005)]{lad05}Lada, C.J., Lada, E.A., 2005 \araa{} 41, 57
\bibitem[Lada et al.(2009)]{lad09} Lada, C.~J., Lombardi, M., \& Alves, J.~F.\ 2009, \apj, 703, 52
\bibitem[Lada et al.(2010)]{lad10} Lada, C.~J., Lombardi, M., \& Alves, J.~F.\ 2010, \apj, 724, 687
\bibitem[Lombardi \& Alves(2001)]{lom01} Lombardi, M., Alves, J., 2001, \aap{} 377, 1023 
\bibitem[Makovoz \& Khan(2005)]{MOPEXCITATION} Makovoz, Khan, 2005, in ASP Conf. Ser. 132, Astronomical Data Analysis Software and Systems VI, ed P.L. Shopbell, M. C. Britton \& R. Ebert (San Francisco: ASP)
\bibitem[Martini(2001)]{mar01} Martini, P., 2001, \aj{} 121,598
\bibitem[Matsuura et al.(2009)]{mat09} Matsuura, M., et al.\ 2009, \mnras{}, 396, 918
\bibitem[McBreen et al.(1979)]{mcb79} McBreen, B., Fazio, G.~G., Stier, M., \& Wright, E.~L.\ 1979, \apjl, 232, L183
\bibitem[McKee(1989)]{mck89} McKee, C.~F.\ 1989, \apj, 345, 782 
\bibitem[Megeath et al.(2009)]{meg09} Megeath, S. T., Allgaier, E., Young, E., Allen, T., Pipher, J. L., Wilson, T.L., 2009, \aj{} 137, 4072
\bibitem[Megeath et al.(2012)]{meg12} Megeath, S. T., et al. 2012 \aj{} 144, 192
\bibitem[Meixner et al.(2006)]{mei06} Meixner, M., et al. 2006 \aj{} 132, 2268
\bibitem[Molinari et al.(2010)]{mol10} Molinari, S., et al., 2010, \aap 518, L100
\bibitem[Motte et al.(2003)]{mot03} Motte, F., Schilke, P., \& Lis, D.C., 2003 \apj{} 582, 277
\bibitem[Motte et al.(2010)]{mot10} Motte, F., et al. 2010, \aap 518, L77
\bibitem[Motte et al.(2012)]{mot12} Motte, F., Bontemps, S., Hennemann, M., et al.\ 2012, SF2A-2012: Proceedings of the Annual meeting of the French Society of Astronomy and Astrophysics, 45
\bibitem[Mu\~{n}oz et al.(2007)]{mun07} Mu\~{n}oz, D., Mardones, D., Garay, G., Rebolledo, D., 2007, \apj{}, 668, 906
\bibitem[Neckel(1978)]{nec78} Neckel, T., 1979, \aap{} 69, 51
\bibitem[Persi \& Tapia(2008)]{per08} Persi, P., Tapia, M., 2008, Handbook of Star Forming Regions Vol II
\bibitem[Peterson et al.(2011)]{pet11gbs3} Peterson, D.~E., Caratti o Garatti, A., Bourke, T.~L., et al.\ 2011, \apjs, 194, 43 
\bibitem[Pinheiro et al.(2010)]{pin10} Pinheiro, M.~C., Copetti, M.~V.~F., \& Oliveira, V.~A.\ 2010, \aap, 521, A26
\bibitem[Pomp{\'e}ia et al.(2008)]{pom08} Pomp{\'e}ia, L., Hill, V., Spite, M., et al.\ 2008, \aap, 480, 379
\bibitem[Probst et al.(2004)]{NFreference} Probst, R.~G., Gaughan, N., Abraham, M., et al.\ 2004, \procspie, 5492, 1716
\bibitem[Rieke et al.(2004)]{rie04} Rieke, G. H., et al., 2004, \apjs{} 154, 25
\bibitem[Robitaille et al.(2006)]{rob06} Robitaille, T.~P., Whitney, B.~A., Indebetouw, R., Wood, K., \& Denzmore, P.\ 2006, \apjs, 167, 256 
\bibitem[Robitaille et al.(2007)]{rob07} Robitaille, T. P., Whitney, B. A., Indebetouw, R., \& Wood, K., 2007, \apjs{} 169, 328
\bibitem[Robitaille et al.(2008)]{rob08} Robitaille, T. P., et al. 2008, \aj{} 136, 2413
\bibitem[Rodriguez et al.(1982)]{rod82} Rodriguez, L.~F., Canto, J., \& Moran, J.~M.\ 1982, \apj, 255, 103 
\bibitem[Romero et al.(2012)]{rom12} Romero, G.~A., Schreiber, M.~R., Cieza, L.~A., et al.\ 2012, \apj, 749, 79
\bibitem[Russeil et al.(2012)]{rus12} Russeil, D., Zavagno, A., Adami, C., et al.\ 2012, \aap, 538, A142
\bibitem[Russeil et al.(2013)]{rus13} Russeil, D., Schneider, N., Anderson, L.~D., et al.\ 2013, \aap, 554, A42 
\bibitem[Rygl et al.(2013)]{ryg13} Rygl, K.L.J., Wyrowski, F., Schuller, F., Menten, K.M., 2013, \aap{}, 549, A5
\bibitem[Schuster et al.(2006)]{sch06} Schuster, M. T., Marengo, M., \& Patten, B. M. 2006, Proc. SPIE, 6270, 65
\bibitem[Skrutskie et al.(2006)]{skr06} Skrutskie, M. F., et al., 2006, \aj{} 131, 1163. 
\bibitem[Spezzi et al.(2011)]{spe11gbs4} Spezzi, L., Vernazza, P., Mer{\'{\i}}n, B., et al.\ 2011, \apj, 730, 65 
\bibitem[Stetson(1987)]{ste87} Stetson, P. B., 1987, \pasp{} 99, 191
\bibitem[Strafella et al.(2010)]{str10} Strafella, F., Elia, D., Campeggio, L., Giannini, T., Lorenzetti, D., Marengo, M., Smith, H. A., Fazio, G. G., De Luca, M., Massi, F., 2010, \apj{} 719, 9
\bibitem[Straw \& Hyland(1989)]{str89} Straw, S. M., Hyland, A. R., 1989, \apj{} 340, 318
\bibitem[Swaters et al.(2009)]{swa11} Swaters, R.~A., Valdes, F., \& Dickinson, M.~E.\ 2009, in Astronomical Data Analysis Software and Systems XVIII, ed. D. A. Bohlender, D. Durand, \& P. Wodlwer (San Francisco, CA: ASP), ASP Conf. Ser., 411, 506
\bibitem[Tapia, Persi, \& Roth(1996)]{tap96} Tapia, M., Persi, P., Roth, M., 1996, \aap{} 316, 102
\bibitem[van Aarle et al.(2011)]{van11} van Aarle, E., van Winckel, H., Lloyd Evans, T., et al.\ 2011, \aap, 530, A90
\bibitem[van Winckel(2003)]{vanw03} van Winckel, H.\ 2003, \araa, 41, 391
\bibitem[Werner et al.(2004)]{wer04} Werner, M. W., et al. 2004, \apjs{} 154, 1
\bibitem[Woods et al.(2011)]{woo11} Woods, P.~M., Oliveira, J.~M., Kemper, F., et al.\ 2011, \mnras, 411, 1597
\bibitem[Zernickel et al.(2013)]{zer13} Zernickel, A., Schilke, P., \& Smith, R.~J.\ 2013, \aap, 554, L2
\bibitem[Zhang et al.(2009)]{zha09} Zhang, Q., Wang, Y., Pillai, T., Rathborne, J., 2009, \apj{} 696, 268
\end{thebibliography}
\end{document}